\documentclass[%
 reprint,              % enable this to get the normal 2 column journal style!
 amsmath,amssymb,
 aps,
prd,
floatfix,
]{revtex4-1}

\usepackage{graphicx}% Include figure files
\usepackage{subcaption}
\usepackage{dcolumn}% Align table columns on decimal point
\usepackage{bm}% bold math
\usepackage{hyperref}% add hypertext capabilities
\usepackage{latexsym}
\usepackage{tabularx}
\newcommand{\be}{\begin{equation}}
\newcommand{\ee}{\end{equation}}
\newcommand{\ba}{\begin{eqnarray}}
\newcommand{\ea}{\end{eqnarray}}
\newcommand{\ban}{\begin{eqnarray*}}
\newcommand{\ean}{\end{eqnarray*}}
\newlength{\subfigheight}

\newcommand{\comment}[1]{}

\graphicspath{{./figures/}}

\begin{document}

%\begin{flushright}
%Sofia University\\
%\end{flushright}
%%%%%%%%%%%%%%%%%%%%%%%%%%%%%%%%%%%%%%%%%%%%%%%%%%%%%%%%%%%%%%%%%%%

\title {Multiple shadows from distorted static black holes}
\author{
Jai Grover$^{1}$\footnote{E-mail:\texttt{jai.grover@esa.int}}, \,Jutta Kunz$^{2}$\footnote{E-mail:\texttt{jutta.kunz@uni-oldenburg.de}},\,\, Petya Nedkova$^{2,3}$\footnote{E-mail: \texttt{pnedkova@phys.uni-sofia.bg}},\\
    Alexander Wittig$^{4}$\footnote{E-mail:\texttt{a.wittig@soton.ac.uk}},\,\, Stoytcho Yazadjiev$^{3,5}$\footnote{E-mail:\texttt{yazad@phys.uni-sofia.bg}}\\
  {\footnotesize${}^{1}$ ESA -- Advanced Concepts and Studies Office, European Space Research Technology Centre} \\
  {\footnotesize Keplerlaan 1, Postbus 299, NL-2200 AG Noordwijk, The Netherlands} \\
  {\footnotesize${}^{2}$  Institut f\"{u}r Physik, Universit\"{a}t Oldenburg}\\
  {\footnotesize D-26111 Oldenburg, Germany}\\
  {\footnotesize${}^{3}$ Faculty of Physics, Sofia University, 5 James}\\
  {\footnotesize   Bourchier Boulevard, Sofia~1164, Bulgaria }\\
  {\footnotesize${}^{4}$ Astronautics Group, University of Southampton} \\
  {\footnotesize Southampton, SO17 1BJ, United Kingdom}\\
   {\footnotesize${}^{5}$ Institute of Mathematics and Informatics}\\
{\footnotesize Bulgarian Academy of Sciences, Acad. G. Bonchev 8} \\
  {\footnotesize  Sofia 1113, Bulgaria}}
\date{\today}

\begin{abstract}
We study the local shadow of the Schwarzschild black hole with a quadrupole distortion and the influence of the external gravitational field on the photon dynamics. The external matter sources modify the light ring structure and lead to the appearance of multiple shadow images. In the case of negative quadrupole moments we identify the most prominent mechanism causing multiple shadow formation. Furthermore, we obtain a condition under which this mechanism can be realized. This condition depends on the quadrupole moment, but also on the position of the observer and the celestial sphere.
\end{abstract}

\maketitle

\section{\label{sec:introduction}Introduction}
Strong gravity tests constitute a major challenge in current gravitational research. The experimental detection of gravitational waves opened a new channel for extracting information about the gravitational interaction in the strong-field regime. At the same time the observational resources in the electromagnetic spectrum are being refined reaching better resolutions, and incorporating further revealing phenomena \cite{Doeleman}-\cite{Zhang}. Both observational channels provide complementary information about strong gravity effects, contributing to the field of multi-messenger astronomy.

One of the most prominent observational projects in the electromagnetic spectrum  is the Event Horizon Telescope \cite{Doeleman}. It represents a global network of radio telescopes, aiming to provide the first image of the supermassive black hole located at the center of our Galaxy. A black hole surrounded by a bright background appears as a dark area, called a black hole shadow. This phenomenon is formed by differentiating between photon trajectories that are captured by the black holes, and those that can escape to infinity and reach the observer. Thus, the black hole shadow represents a gravitationally lensed image of the photon region. Its appearance is usually associated with the presence of an event horizon, however not exclusively, since it can be observed by more exotic horizonless compact objects such as wormholes and gravastars \cite{Nedkova:2013}-\cite{Sakai}.

Black hole shadows were first studied theoretically in the context of the Kerr-Newman family in the classical work of Bardeen \cite{Bardeen}. Recently, their properties were investigated for further black hole solutions in general relativity \cite{Grenzebach:2014}-\cite{Tinchev:2014}, and in various modified theories of gravity \cite{Amarilla:2010}-\cite{Yazadjiev:2018}. The shape and the particular deformation of the black hole shadow carries information about the properties of the compact object, such as its spin and multipole moments \cite{Tsukamoto}-\cite{Li}. Supplemented by further information from observations in the electromagnetic spectrum, such as the pulsar timing for example, it can be used for testing the Kerr hypothesis, or differentiating between alternative gravitational theories \cite{Johannsen}-\cite{Bambi:2015}.

Particularly interesting are cases, when the black hole does not occur in isolation, but interacts with another gravitational source. Such scenarios provide different approximations of the realistic astrophysical black hole settings in the centre of galaxies, or as a part of binary systems. Simple models include static double black hole configurations in equilibrium \cite{Yumoto}-\cite{Shipley},  or black holes interacting with a scalar field condensate \cite{Cunha:2015}-\cite{Cunha:2016}. Recently, the fully dynamical black hole merger was studied by obtaining the observable image of two coalescent black holes \cite{Bohn}.

Another possibility for investigating the shadow of black holes influenced by an external gravitational field is the class of distorted black holes \cite{Geroch}. They represent local exact solutions describing the near-horizon region of black holes interacting with a quasi-stationary matter distribution such as an accretion disk, or a binary companion in the very initial stages of the inspiral. In general the distorted black holes are characterized by an infinite number of parameters connected with the interior multipole moments of the external matter distribution. In this way they encode information about the impact of the surrounding gravitational sources on the central black hole.

As a local solution the distorted black hole solution is valid only in a certain finite region encompassing the horizon. Its precise extent is model specific, and depends on the particular black hole and matter system, which is described. A global solution is constructed by matching the interior distorted black hole solution with an exterior solution containing only the external matter sources, but not the central object.

Distorted black holes allow us to study the modification of the spacetime properties when the black hole is not considered in isolation in terms of exact solutions. Hence, they are particularly valuable for describing phenomena in the strong field regime, where approximate or perturbative solutions do not provide an adequate description, while at the same time giving better intuition than numerical solutions. Experimental tests probing strong gravity physics in the black hole vicinity rely crucially on the particular solutions to the gravitational field equations used to describe the spacetime. In this respect distorted black holes can provide valuable intuition for modelling and interpreting observational data.

Several distorted black hole solutions were constructed in the literature generalizing existing isolated black holes. An early example was obtained by Doroshkevich, Zeldovich and Novikov \cite{Doroshkevich}, describing a Schwarzschild black hole in an external quadrupole gravitational field. Geroch and Hartle \cite{Geroch} studied the most general static vacuum distorted black hole with a regular event horizon in classical general relativity, while the explicit form of the corresponding metric was obtained completely in \cite{Breton:1997}. Rotating and charged solutions were constructed in \cite{Tomimatsu}-\cite{Breton:1998}, which were further extended to higher dimensions \cite{Abdolrahimi:2010}-\cite{Nedkova:2014}. Recently, a class of magnetized distorted black holes was obtained, taking into account also  the influence of the magnetic field generated by an external source \cite{Kunz:2017}.

In a number of works certain physical properties of distorted black holes were investigated, such as  the ergoregion configurations for the distorted Kerr black hole \cite{Abdolrahimi:2015a}, the geodesic motion in the equatorial plane, and the stability of the photon region for the distorted Schwarzschild black hole \cite{Shoom:2016}-\cite{Shoom:2017}. The local shadow cast by the Schwarzschild black hole in an external quadrupole and octupole gravitational fields was studied in \cite{Abdolrahimi:2015b}-\cite{Abdolrahimi:2015c}, where  multiple shadow structures were observed in the case of octupole distortion.

The purpose of our work is to study in detail the light ring structure and shadow of the Schwarzschild black hole with a quadrupole distortion, and shed some light on the dynamical mechanisms leading to shadow properties. Light rings, or equivalently planar circular photon orbits, are of great  interest  for the dynamics of the geodesic motion, as they represent fixed points of the underlying dynamical system, and are closely related to observable features. Unstable light rings are associated with the shape of the shadow \cite{Grover}, while their orbital frequency and instability timescale are related to the quasi-normal modes spectrum of the black hole \cite{Cardoso:2009}. The stable ones, on the other hand,  lead to chaotic scattering in the lensing and spacetime instabilities \cite{Shipley}, \cite{Cunha:2016}, \cite{Dolan:2016}, \cite{Cunha:2017}. In our work we classify completely the possible light ring configurations for the Schwarzschild black hole with a quadrupole distortion. In addition to the circular photon orbits lying at the equatorial plane, we discover that off-equatorial light rings exist located in two horizontal planes, which are symmetric with respect to the equatorial one.

The distorted Schwarzschild solution is a local solution, hence we can consider only a local shadow, i.e. a shadow observed at some finite distance from the black hole within the validity of the distorted solution, and not at asymptotic infinity. This introduces further parameters into the problem - the radial position of the observer, as well as that of the celestial sphere, on which we assume all light sources to reside. The position of the celestial sphere can be associated with the region of validity of the local back hole solution as it defines a radial cut-off for the lensing analysis. However, we cannot define an absolute upper bound for its allowed values, since this region is model dependent. In every specific case it is determined by the particular black hole and matter system we want to describe. 

Our analysis demonstrates that a multiple shadow hierarchy is also present in the case of quadrupole distortion, which was not reported in previous work \cite{Abdolrahimi:2015b}-\cite{Abdolrahimi:2015c}. Moreover, in the case of a negative quadrupole moment, we are able to define a sufficient condition for the appearance of multiple shadows. This condition depends not only on the value of the quadrupole moment, but also on the position of the observer and the celestial sphere, such that, for any negative value of the quadrupole moment, there exists a distant observer who will see multiple shadows. Hence, the appearance of multiple shadows is a qualitative effect caused by the mere presence of distortion by an external gravitational source, and not by its strength. It can be switched off only by removing the external matter distribution ending up with an isolated black hole.

Based on this, we conjecture that in general the appearance of a multiple shadow structure is a manifestation of the interaction of the black hole with a second source of gravitational field. In the literature we have examples where the additional gravitational source is a second black hole \cite{Yumoto}-\cite{Shipley}, or a compact distribution of a scalar field condensate in the form of a torus surrounding the black hole \cite{Cunha:2015}-\cite{Cunha:2016}. In our work we demonstrate that the same effect occurs due to the interaction with a rather general matter distribution with a quadrupole symmetry.

The paper is organized as follows. In the next section we present the metric of the distorted Schwarzschild solution, and discuss its relevant properties, such as regularity conditions, and horizon deformation. In section III we present the Hamiltonian formulation of the geodesic motion, which we use in our investigations, and introduce a 2-dimensional effective potential. The effective potential defines the allowed and forbidden regions for photon motion, thus determining its qualitative behavior. In section IV we describe the lensing set-up we use to obtain our images. Section V is devoted to the analysis of the possible light ring configurations. Three types of qualitatively different light ring configurations are identified, depending on the value of the quadrupole moment. One of them contains a light ring pair outside the equatorial plane. The behavior of the light ring positions is analyzed when varying the quadrupole moment, and their influence on the photon dynamics. In section VI we develop an argument explaining the appearance of multiple shadows in the case of a negative quadrupole moment. We provide a mechanism for generating multiple images in the shadow based on scattering from a forbidden region, which is formed due to the presence of an external gravitational source. For its realization certain conditions are needed, defining a threshold, which depends simultaneously on the value of the quadrupole moment, the position of the observer and the celestial sphere.

%%%%%%%%%%%%%%%%%%%%%%%%%%%%%%%%%%%%%%%%%%%%%%%%%%%%%%%%%%%%%%%%%%%%%%%%%%%%

\section{\label{sec:solution}Distorted Schwarzschild black hole}
The metric of the distorted Schwarzschild black hole is given in the form \cite{Breton:1997}

\begin{eqnarray}\label{dist_Sch}
ds^2 &=& -\frac{x-1}{x+1}e^{2 U} dt^2 +  \sigma^2(x+1)^2(1-y^2)e^{-2 U}d\varphi^2 \nonumber \\[2mm]
&+&  \sigma^2(x+1)^2e^{2\gamma-2U} \left(\frac{dx^2}{x^2-1}+\frac{dy^2}{1-y^2}\right), \nonumber \\[2mm]
U &=& \sum^{\infty}_{n=0}a_nR^n P_n\left(\frac{xy}{R}\right), \quad ~~~ R = \sqrt{x^2 +  y^2 - 1}, \nonumber \\[2mm]
\gamma &=& \sum^{\infty}_{n,k=1}\frac{nk}{n+k}a_na_k R^{n+k}\left(P_n P_k - P_{n-1}P_{k-1}\right) \nonumber \\
&+& \sum^{\infty}_{n=1}a_n\sum^{n-1}_{k=0}\left[(-1)^{n-k+1}(x+y)\right. \nonumber \\[2mm]
&-&\left.x+y\right]R^k P_k\left(\frac{xy}{R}\right), 
\end{eqnarray}
where $P_n(\frac{xy}{R})$ are the Legendre polynomials. The prolate spheroidal coordinates $(x,y)$ take the ranges $x\geq 1$, and $-1\leq y\leq 1$ in the domain of outer communication, as the physical infinity is located at $x\rightarrow\infty$. The solution contains a Killing horizon located at $x=1$, and the symmetry axis consists of two disconnected components  at $y=1$ and $y=-1$.

The metric function $U$ is a harmonic function in an auxiliary non-physical flat space, while  $\gamma$ is a solution to the linear system

\begin{eqnarray}\label{gamma}
\partial_x\gamma &=& \frac{1-y^2}{x^2-y^2}\left[x(x^2-1)(\partial_x U)^2 - x(1-y^2)(\partial_y U)^2\right. \nonumber  \\[1mm]
&-& \left. 2y(x^2-1)\partial_x U\,\partial_y U  
+  2x\partial_xU - 2y\partial_yU\right], \nonumber \\[2mm]
\partial_y\gamma &=& \frac{x^2-1}{x^2-y^2}\left[y(x^2-1)(\partial_x U)^2
- y(1-y^2)(\partial_y U)^2 \right.  \nonumber \\[1mm]
&+& \left. 2x(1-y^2)\partial_x U\,\partial_y U\right] \nonumber \\[1mm]
&+& \frac{1}{x^2-y^2}\left[2y(x^2 - 1)\partial_xU + 2x(1 - y^2)\partial_yU\right].
\end{eqnarray}
These functions are determined by a discrete set of real constants $a_n, \, n \in\mathcal{N}$, and lead to a deviation from asymptotic flatness if the distorted Schwarzschild black hole is considered as a global solution. It is physically more relevant to regard the metric ($\ref{dist_Sch}$) as  a local solution valid in some neighbourhood of the horizon. Then, $U$ and $\gamma$ are interpreted as arising due to interaction with a gravitational source located in the exterior of the region of validity. The parameters $a_n, \, n \in\mathcal{N}$  characterize the external gravitational potential, corresponding to the interior multipole moments in its expansion. In the limit when all the constants $a_n$ vanish, the isolated Schwarzschild black hole is recovered. The solution possesses a further real parameter $\sigma$, equal to the Komar mass on the black hole horizon. It coincides with the Komar mass of the isolated Schwarzschild black hole, which is not influenced by the external field.

For balanced solutions the external matter is restricted by the condition for absence of conical singularities on the axis. It reduces to the following constraint on the solution parameters

\begin{eqnarray}\label{balance_Sch}
\sum^{\infty}_{n=0}a_{2n+1} =0.
\end{eqnarray}

Another restriction is introduced by the requirement that the material sources, which generate the external gravitational potential, should satisfy the strong energy condition. This leads to the condition $U\leq 0$ \cite{Geroch}, implying that

\begin{eqnarray}\label{SEC}
u_0 = \sum^{\infty}_{n=0}a_{2n} \leq 0.
\end{eqnarray}

The interaction with the external potential leads to a deformation of the horizon geometry with respect to the spherical one. The metric on the horizon cross-section is given by

\begin{eqnarray}\label{hor_metricDS}
ds_H^2 =  4\sigma^2e^{\gamma-2U}\left[e^{-\gamma(y)}(1-y^2)d\varphi^2
+ e^{\gamma(y)}\frac{dy^2}{1-y^2}\right] . \nonumber
\end{eqnarray}
Since the combination of the metric functions $U - \frac{1}{2}\gamma$ reduces to a constant on the horizon, the horizon geometry is determined by a single function $U(y)$. It encodes the deviation from the geometrical sphere, hence it can be regarded as a shape function \cite{Hajicek:1973}. The horizon area takes the form

\begin{eqnarray}
A_H &=& 16\pi\sigma^2e^{-2U}|_{x=1, y=\pm 1} = 4\pi R_H^2, \\ \nonumber
R_H &=& 2\sigma\exp{\left[-\sum^{\infty}_{n=0}a_{2n}\right]},
\end{eqnarray}
and the scalar quantity $R_H$, appearing also as a scale factor in ($\ref{hor_metricDS}$), is interpreted as an effective horizon radius in analogy with the spherical case.

For the investigation of geodesic motion in the distorted Schwarzschild spacetime it is convenient to perform a suitable rescaling of the metric
\begin{eqnarray}
ds^2 \rightarrow \Omega^{-2}ds^2, \quad~~~~~ \Omega^2 = \sigma^2 e^{-2u_0}.
\end{eqnarray}
\noindent
The rescaled metric becomes

\begin{eqnarray}\label{dist_Sch1}
ds^2 &=& -\frac{x-1}{x+1}e^{2 \widetilde U} dt^2 + (x+1)^2(1-y^2)e^{-2 \widetilde U}d\varphi^2 \nonumber \\[2mm]
&+&  (x+1)^2e^{2\gamma-2\widetilde U} \left(\frac{dx^2}{x^2-1}+\frac{dy^2}{1-y^2}\right), \\[2mm]
\widetilde U &=& U -u_0,
\end{eqnarray}
where we have redefined the time coordinate as

\begin{eqnarray}
t\rightarrow \sigma^{-1}e^{2u_0} t.
\end{eqnarray}
It describes a solution with a unit mass, and a horizon area $A_H = 16\pi^2$, which is independent of the external field, and coincides with that of the corresponding isolated Schwarzschild black hole.
We can  represent the solution in the conventional Schwarzschild coordinates $(r, \theta)$ by performing the transformation
\begin{eqnarray}
x =r-1, \quad~~~ y = \cos\theta.
\end{eqnarray}

Distorted black holes are most generally divided into solutions with even and odd distortion by setting only the even or odd multipole moments $a_n$ different from zero. Each class of solutions corresponds to different symmetries of the external gravitational source. Physically most relevant are the cases when only the lowest even or odd multipole moments do not vanish. They would describe the interaction with an external gravitational field, which is considerably weaker than that of the central object. The solution with the lowest even distortion is the quadrupole case, when  $a_2\neq 0$, and all the rest of the multipole moments vanish. Due to the regularity condition ($\ref{balance_Sch}$), the simplest solution with an odd distortion is the octupole case, when  $a_3 = -a_1\neq 0$, and all the rest of the multipole moments vanish.

In our work we investigate the light propagation for spacetimes with quadrupole distortion. In this case the metric functions are given by

\begin{eqnarray}\label{U_quad}
{\widetilde U}&=&\frac{a_2}{2}(3x^2y^2-x^2-y^2-1), \\[1mm]
\gamma&=&2a_2x(y^2-1)+\frac{1}{4}a_2^2\left[(x^2+y^2-1)\times\right. \nonumber\\[1mm]
&&\left.(x^2+y^2-10x^2y^2) -x^2-y^2+9x^4y^4+1\right]. \nonumber
\end{eqnarray}

%%%%%%%%%%%%%%%%%%%%%%%%%%%%%%%%%%%%%%%%%%%%%%%%%%%%%%%%%%%%%%%%%%%%%%%%%%%%

\section{\label{sec:geodesicequations} Geodesic equations}
We consider the Hamiltonian formulation of the geodesic motion. Free particles are described by the Hamiltonian

\begin{equation}
H = \frac{1}{2}g^{\mu \nu} p_\mu p_\nu ,
\label{ham}
\end{equation}
where $p_\mu$ is the canonical momentum, and the Hamilton equations are given by
\begin{equation}
{\dot q^\mu} =\frac{\partial H}{\partial p_\mu} \quad , \quad
{\dot p_\mu}=-\frac{\partial H}{\partial q^\mu}.
\end{equation}
We denote by an overdot differentiation with respect to an affine parameter along the geodesic, and $q^\mu$ are the spacetime coordinates. For stationary and axisymmetric spacetimes two conserved quantities exist
\begin{eqnarray}
p_t = - E, \quad~~~ p_\varphi = L,
\end{eqnarray}
representing the particle's energy and its angular momentum with respect to the symmetry axis.
In the case of the distorted Schwarzschild solution the canonical momentum takes the form
\begin{eqnarray}
\label{momenta}
  p_t &=& -\frac{x-1}{x+1}e^{2\widetilde U}\, \dot t = -E, \label{e} \\[2mm]
  p_\varphi &=& (x+1)^2(1-y^2)e^{-2\widetilde U} \dot\varphi = L,  \\[2mm] \nonumber
  p_x &=& \frac{x+1}{x-1}e^{2\gamma -2\widetilde U}\, \dot x, \\[2mm] \nonumber
  p_y &=& \frac{(x+1)^2}{1-y^2}e^{2\gamma -2\widetilde U}\, \dot y,  \nonumber
\end{eqnarray}
and we can express the Hamiltonian as
\begin{eqnarray}
H(p,q) &=&  \frac{(x-1)e^{2\widetilde U - 2\gamma}}{2(x+1)}\,p_x^2+\frac{(1-y^2)e^{2\widetilde U - 2\gamma}}{2(x+1)^2}\,p_y^2 \nonumber \\[3mm]
&+& V(x,y,E,L) , 
\end{eqnarray}
\begin{eqnarray}
V(x,y,E,L) = -\frac{(x+1)e^{-2\widetilde U}E^2}{2(x-1)} + \frac{e^{2\widetilde U}L^2}{2(x+1)^2(1-y^2)}. \nonumber
\end{eqnarray}
\noindent
The function $V$ depends only on the spacetime coordinates and the constants of motion. Hence, it can be interpreted as a 2-dimensional effective potential governing the particle motion, while the remaining part of the Hamiltonian corresponds to a kinetic term $T = \frac{1}{2}g^{ij}p_ip_j$, $i,j = x, y$. The Hamiltonian vanishes identically on null geodesics, and the kinetic term is always positive. Therefore, the allowed regions for photon motion are defined by the condition $V\leq 0$. Introducing the impact parameter $\eta = L/E$, the effective potential can be written as
%\begin{eqnarray}
%&&V(x,y,E,L) = -\frac{E^2e^{2\widetilde U}}{2(x+1)^2(1-y^2)}\left(h^2(x,y) %-\eta^2\right),   \nonumber \\[2mm]
%&&h^2(x,y) = \frac{(x+1)^3(1-y^2)}{x-1}e^{-4\widetilde U}.\\  \nonumber
%\end{eqnarray}
\begin{eqnarray}
&&V = -\frac{E^2e^{2\widetilde U}}{2(x+1)^2(1-y^2)}
\left[h(x,y) +\eta\right]\left[h(x,y) -\eta\right],   \nonumber \\[3mm]
&&h(x,y) = +\sqrt{\frac{(x+1)^3(1-y^2)}{x-1}e^{-4\widetilde U}}. 
\end{eqnarray}
\noindent
Consequently, null geodesics with an impact parameter $\eta$ are confined to regions where $h(x,y)\geq|\eta|$ is satisfied.

The qualitative behavior of the effective potential $V$ depends on the multipole moments $a_n$ of the solution, as well as on the absolute value of a geodesic's impact parameter $|\eta|$, since the problem is symmetric with respect to the shift $\eta\rightarrow -\eta$. In the case of quadrupole distortion, when the metric function $\widetilde U$ is given by ($\ref{U_quad}$), the effective potential is analyzed in detail in Section \ref{sec:photonorbits}. Although different values of $|\eta|$ lead to a range of qualitatively different configurations, the behavior of the potential at large distances is determined only by the sign of the quadrupole moment. Analyzing the asymptotic expansion of the potential we obtain that the following situations are realized. For negative quadrupole moments there always exists a forbidden region encompassing the equatorial plane for large enough values of $x$. At the same time allowed regions are present in the vicinity of the axes $y=\pm 1$ forming escape channels. For positive quadrupole moments the opposite situation is observed. After a certain value of $x$ the region around the equatorial plane is allowed, while forbidden regions exist in the vicinity of the axes, expanding when the radial distance is increased. In both cases photons with an impact parameter $\eta =0$ are an exception, since then no forbidden regions exist. We further apply this qualitative analysis in building the argument for the observation of multiple shadows in Section
\ref{sec:multipleshadows}.

%%%%%%%%%%%%%%%%%%%%%%%%%%%%%%%%%%%%%%%%%%%%%%%%%%%%%%%%%%%%%%%%%%%%%%%%%%%%

\section{\label{sec:lensingimages} Lensing images}
\begin{figure}[tb]
    \centering
    \includegraphics[width=0.4\textwidth]{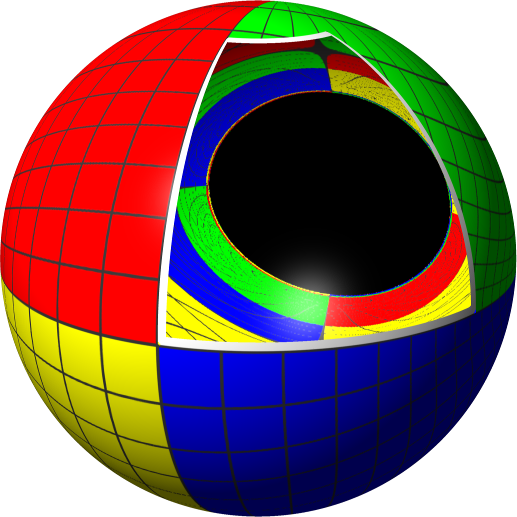}
    \caption{\label{fig:sphere}Illustration of the spherical observer sky for a given spacetime (with partial cutout). The observer is located at the center of this sphere.}
\end{figure}

\begin{figure*}[tb]
    \centering
    \begin{subfigure}[t]{0.49\textwidth}
        \includegraphics[width=\textwidth]{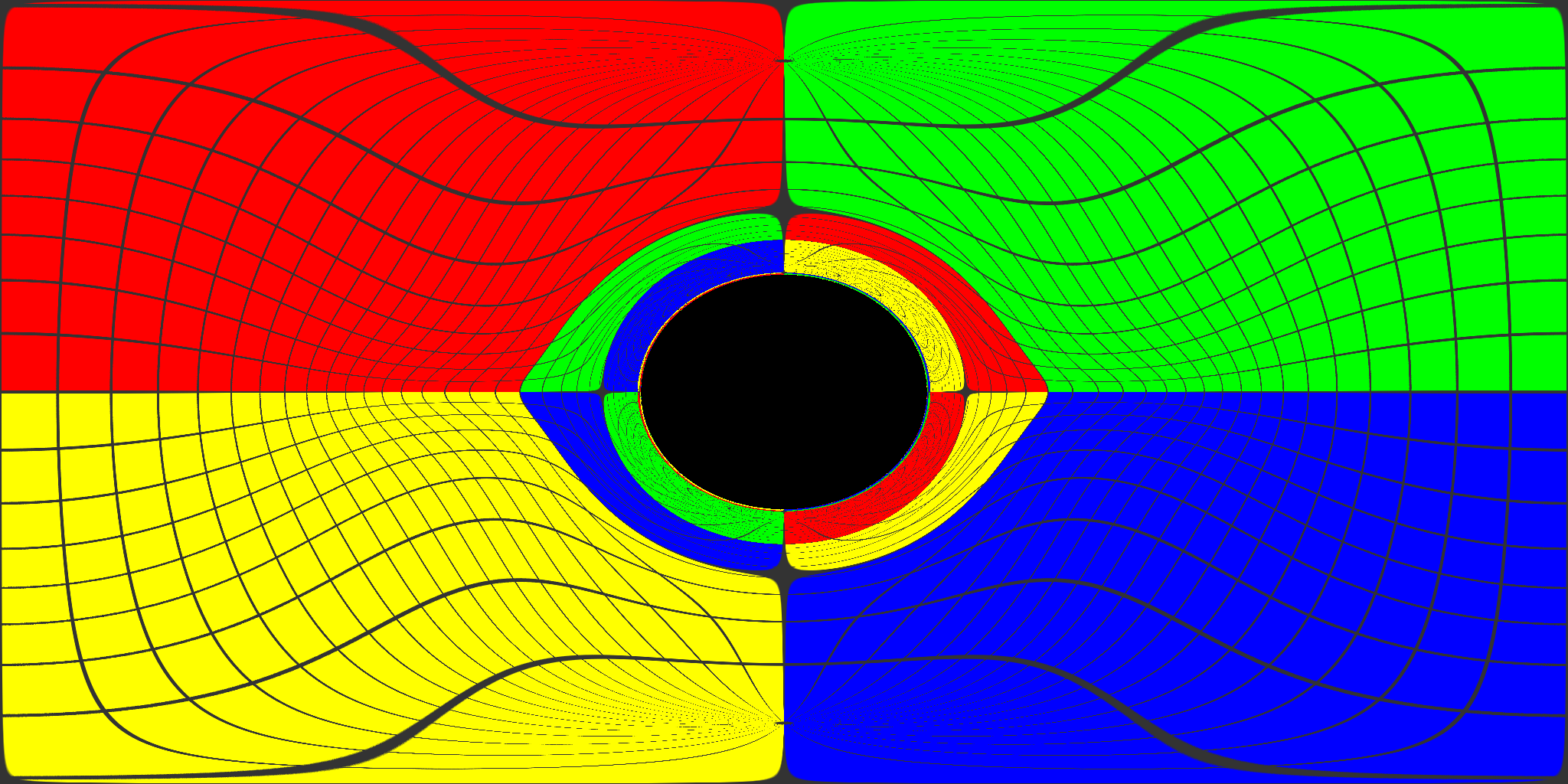}
        \caption{A lensing image in equirectangular projection.}
    \end{subfigure}
    \begin{subfigure}[t]{0.49\textwidth}
        \includegraphics[width=\textwidth,trim={81mm 45.5mm 94.5mm 48.3mm},clip]{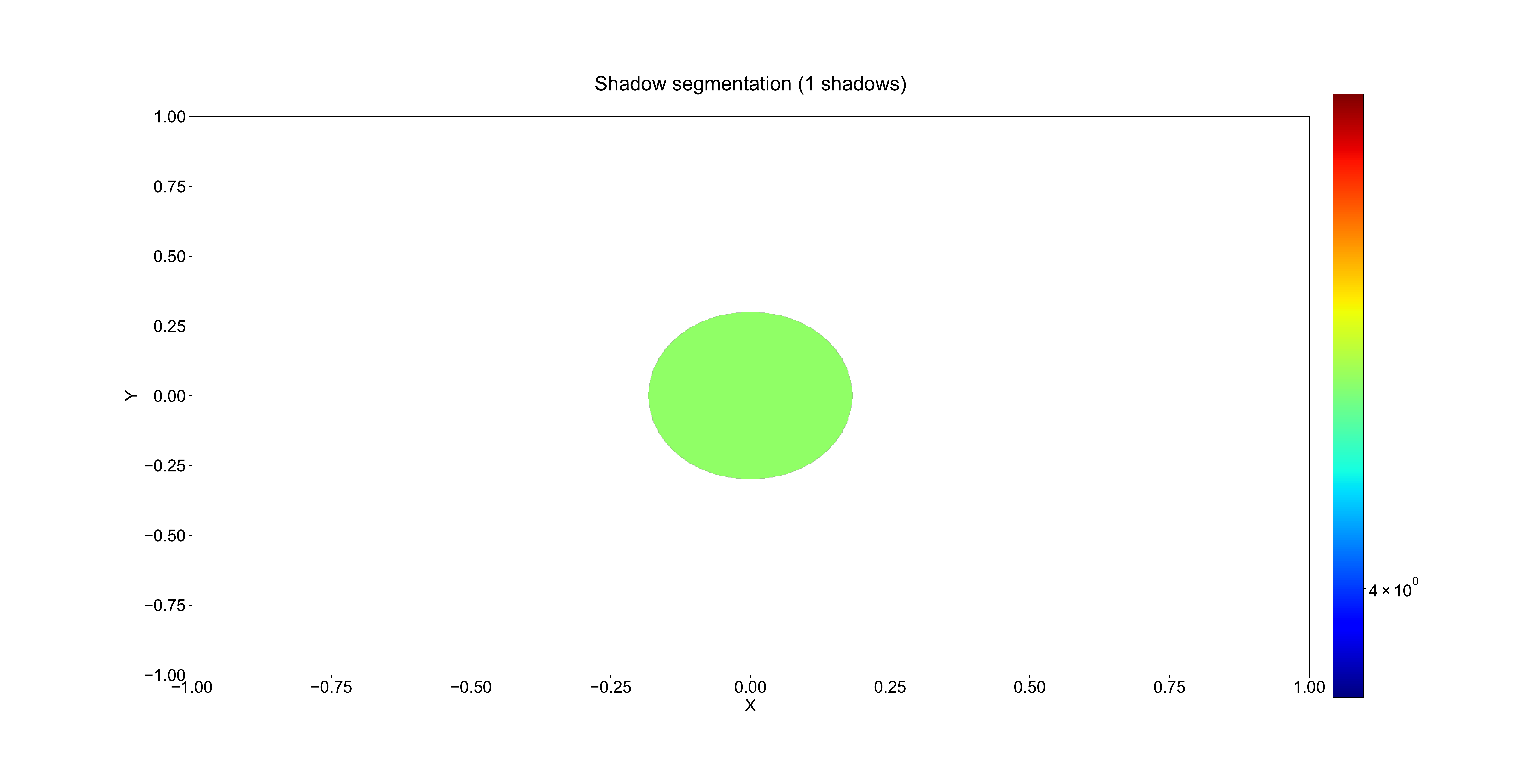}
        \caption{The corresponding shadow image.}
    \end{subfigure}

    \caption{\label{fig:equirect}Illustration of the equirectangular projection for lens and shadow maps for the observer sky shown in Fig. \ref{fig:sphere}.}
\end{figure*}

In our analysis we will make extensive use of lensing images as a tool to study the properties of trajectories following the geodesic equations. To construct such an image we first follow a ray-tracing procedure, integrating null geodesics backwards from points on the observer's local sky to points on a distant celestial sphere, wherein all light sources are assumed to reside \cite{Bohn}. The resulting lens map (Fig. \ref{fig:sphere}) is then projected into a planar image Fig. \ref{fig:equirect}(a). 

In particular, we use an equirectangular projection, which projects a 2-sphere parameterized by the angles $\varphi\in[0,2\pi]$ and $\vartheta\in[0,\pi]$ into the plane by linearly mapping each value of $\varphi$ to $x$ and each $\vartheta$ to $y$. We shall refer to the equirectangular projection of the lens map over all possible viewing angles for a given observer as the \emph{equirectangular full lensing image}, or just \emph{lensing image} for short. For values near $\vartheta=\pi/2$, this projection exhibits little distortion of the resulting image. The equirectangular projection in this region is essentially the linearization of the projection onto the cylinder tangent at the equator. For this reason, we place the black hole in the center of the image, at $\varphi=\pi$ and $\vartheta=\pi/2$. The immediate surroundings of the black hole thus remain relatively undistorted, allowing for uncomplicated analysis. 

Note, however, that regions closer to the poles, corresponding to the local observer's up and down, become very distorted as even small features near the poles get enlarged (see for example the grid lines in Fig. \ref{fig:equirect}(a)). While this trivially follows from the singularity of spherical coordinates at the poles, it is important to keep this in mind when interpreting features seen in the lensing maps. Furthermore, in order for the image to show the black hole shadow at the center while matching our intuitive understanding of left/right and up/down, both axes are flipped. That is, the up direction in the image corresponds to decreasing $\vartheta$ while the right direction corresponds to decreasing $\varphi$.

% Alex: Someone needs to read and improve this for me:
In Fig. \ref{fig:equirect}(b) the shadow image for the same spacetime and observer as Fig. \ref{fig:equirect}(a) is displayed. In this representation only those light rays that pass behind the horizon, defining the black hole's shadow, are drawn. In the example shown, there is only a single connected component corresponding to the central shadow of the (slightly) distorted black hole. In our shadow plots, connected components of the shadow are colored by their size, measured in image pixels. Note that here connectedness is determined based on the pixels in the image, so if the features of a shadow are finer than the size of a single pixel, components that really are connected can be incorrectly identified as non-connected, and vice-versa. This representation is particularly useful when studying the emergence of multiple shadows, as we do in Section \ref{sec:multipleshadows}.

% Jai: maybe we should add some explanation of the projection here so people use it more intuitively to avoid questions like "there's such a big area on the top and bottom" or things like that. Maybe put a 3D panorama of something just to give intuition?
% Alex: I took a stab at it, please modify as needed. Also check out figures/equirect/evil_eye_cut.gif for use in presentations on this stuff ;)

%%%%%%%%%%%%%%%%%%%%%%%%%%%%%%%%%%%%%%%%%%%%%%%%%%%%%%%%%%%%%%%%%%%%%%%%%%%%

\section{\label{sec:photonorbits} Circular photon orbits}

In this section we consider the circular null geodesics, which are also called light rings in the literature,  for black holes with quadrupole distortion ($a_2 \neq 0, a_n =0, n\neq2$). These solutions are characterized with reflection symmetry with respect to the equatorial plane. Therefore, we can distinguish between two types of light rings - such lying in the equatorial plane $y=0$, and such lying in a symmetric pattern in two other planes $\pm y=const$. Geodesic motion in the equatorial plane was also discussed previously in \cite{Shoom:2016}. In the following we describe all the possible light ring configurations, and the ranges of the quadrupole moment $a_2$, for which they are realized.

\subsection{Light rings in the equatorial plane}
Light rings in the equatorial plane are determined by the equations $p_x = p_y =0$, $\dot{p_x}= \dot{p_y}=0$, and  $y=0$. Consequently, they correspond to the stationary points of the effective potential $V(x,0)$, in which it vanishes, or, equivalently, to the stationary points of the potential $h^2(x,0)$. For quadrupole distortion the potential $h^2(x,y)$ is given by

\begin{eqnarray}
h^2(x,y) &=& \frac{(x+1)^3(1-y^2)}{x-1}e^{-4\widetilde U}, \\ \nonumber
\widetilde U &=& \frac{1}{2}a_2(3x^2y^2-x^2-y^2-1).
\end{eqnarray}
Its  stationary points in the equatorial plane satisfy the condition $\partial_x h^2(x,0) = 0$, which reduces to the equation 

\begin{eqnarray}\label {LR0}
x-2 + 2a_2\,x(x^2 -1) = 0.
\end{eqnarray}
It determines the position of the light rings as a function of the quadrupole moment $a_2$. Examining this equation it follows that no stationary points exist for values of the quadrupole moment below a critical value $a_2^{crit} \approx -0.021$. For values in the range $a_2^{crit}<a_2<0$ two stationary points occur, corresponding to  a minimum and a maximum of the potential $h^2(x,0)$, while for $a_2\geq0$ only a minimum is observed. Decreasing the quadrupole moment in the range $a_2^{crit}<a_2<0$ causes the two stationary points to approach each other, until at $a_2=a_2^{crit}$ they merge into an inflection point of $h^2(x,0)$. The position of the inflection point is a solution to the equation

\begin{eqnarray}
x^3 - 3x^2 + 1 = 0.
\end{eqnarray}
It possesses a single real solution in the interval $x\in (1,+\infty)$, which takes the approximate value $x_{crit} \approx 2.879$.  Increasing the quadrupole moment in the range $a_2\geq0$ leads to moving the light ring closer to the horizon.

\begin{figure*}[tb]
    \centering
        \begin{subfigure}[t]{0.32\textwidth}
        \includegraphics[width=\textwidth]{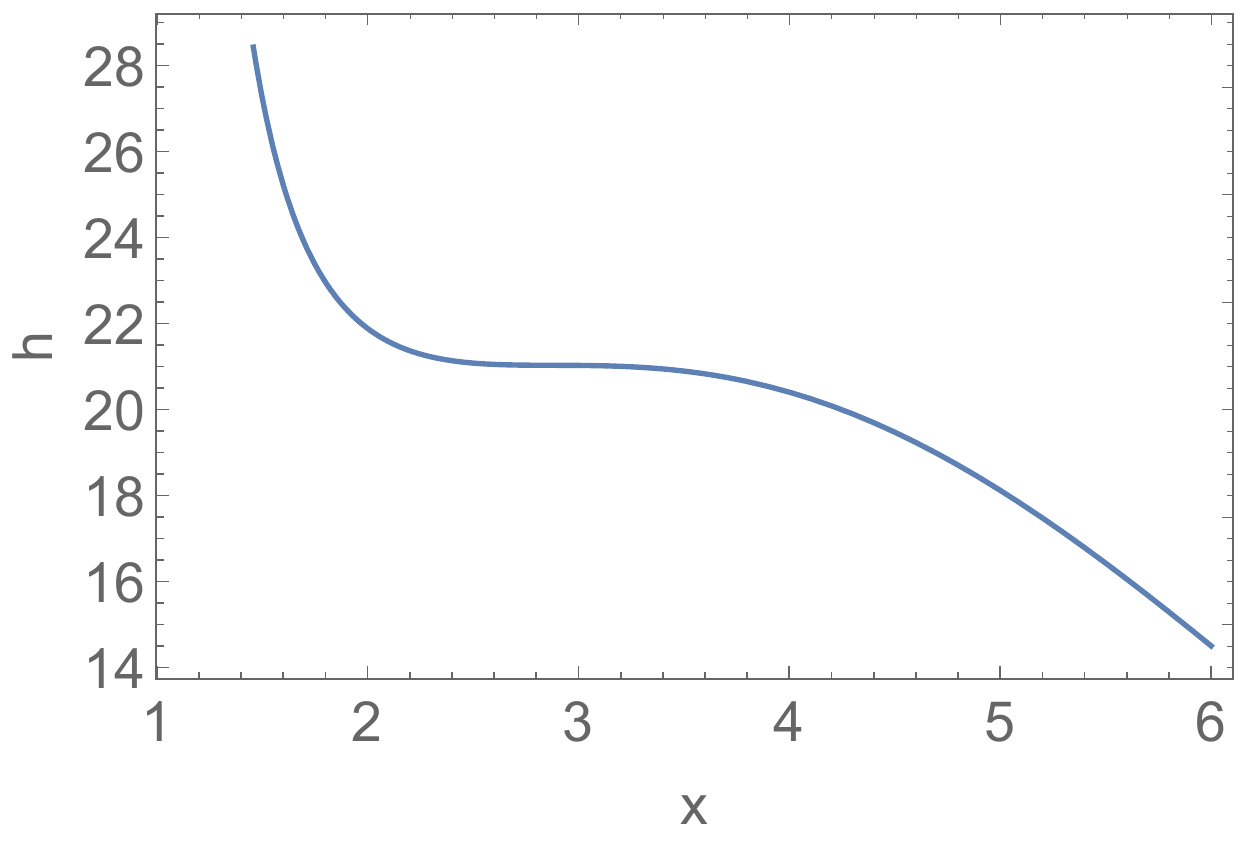}
        \caption{$a_2=-0.021$}
    \end{subfigure}
    %\hspace{1cm}
    \begin{subfigure}[t]{0.32\textwidth}
        \includegraphics[width=\textwidth]{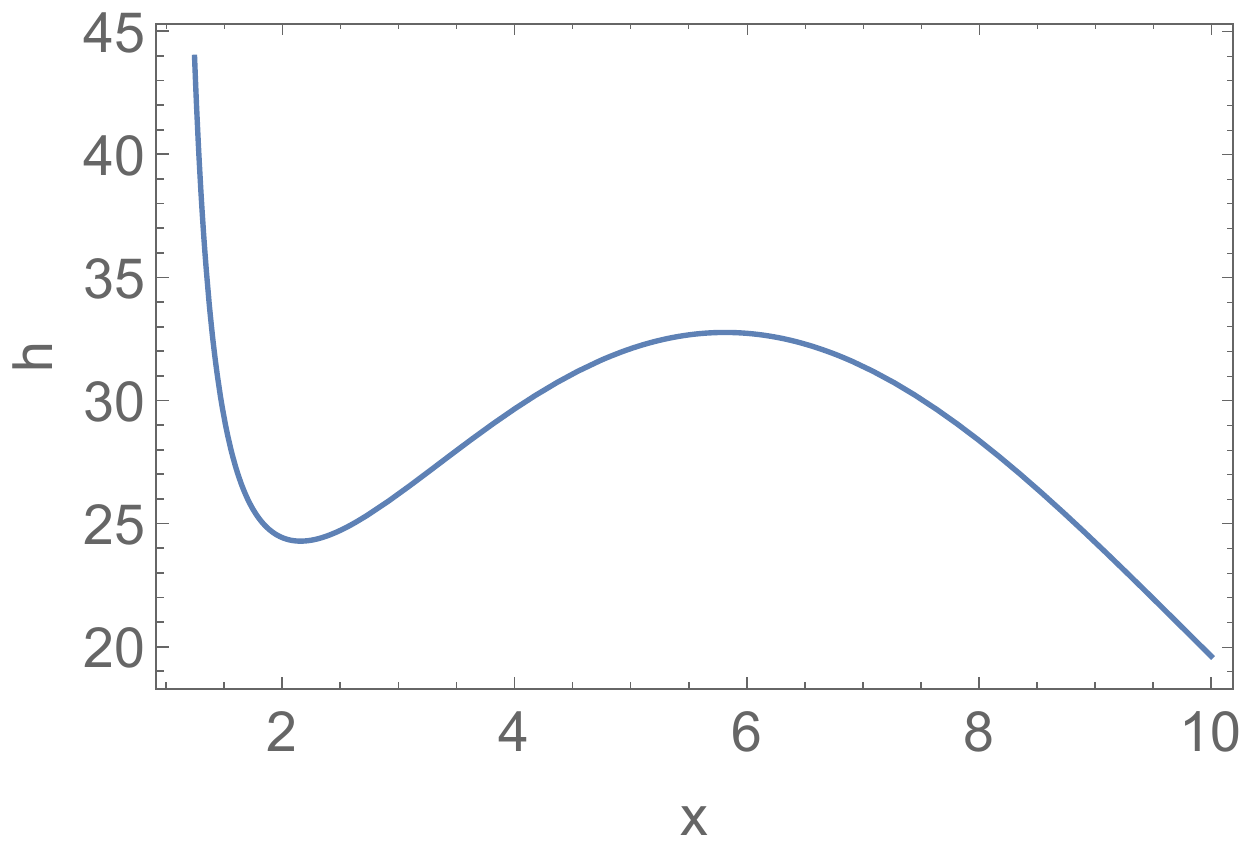}
        \caption{$a_2=-0.01$}
    \end{subfigure}
        \begin{subfigure}[t]{0.32\textwidth}
        \includegraphics[width=\textwidth]{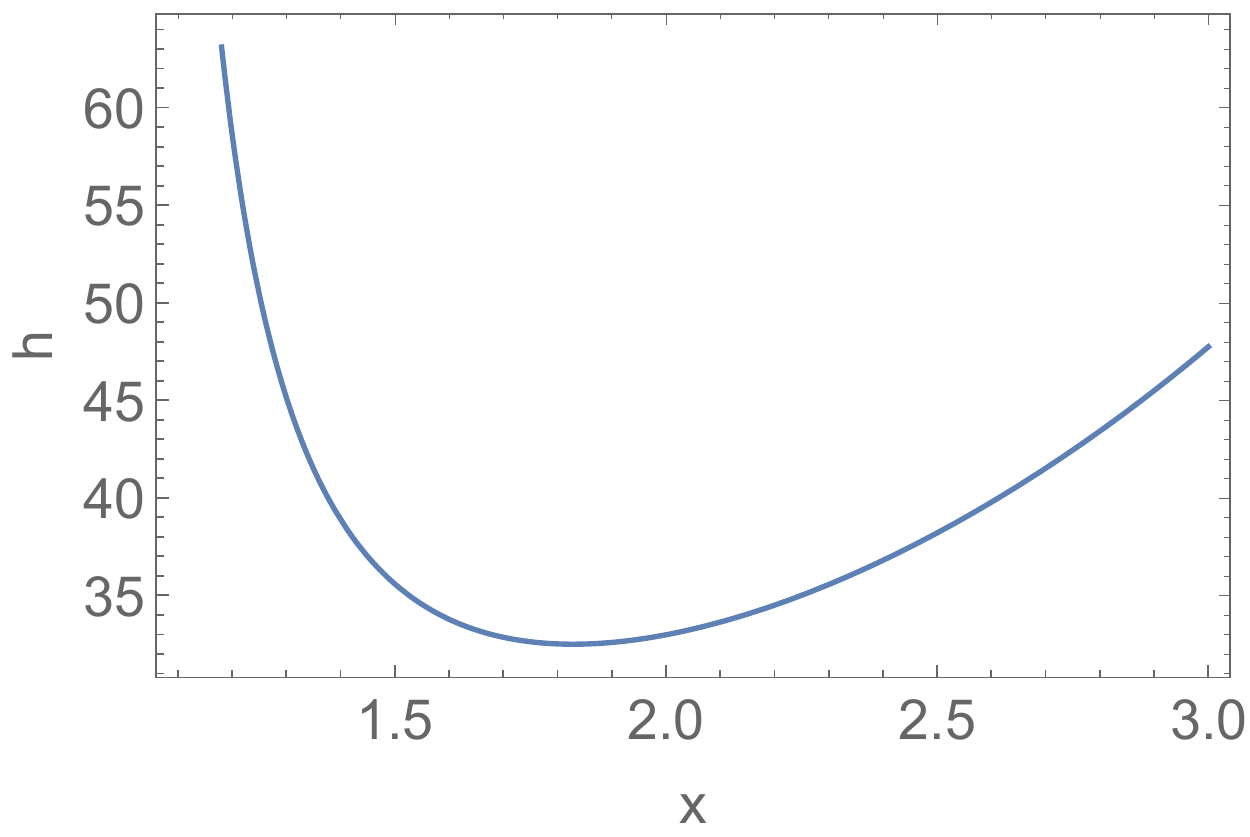}
        \caption{$a_2=0.02$}
    \end{subfigure}
        \caption{\label{h0}The effective potential $h^2(x,0)$ in the equatorial plane for different values of the quadrupole moment: a) For $a_2<a_2^{crit} \approx -0.021$  no stationary points exist, and consequently no light rings. At $a_2^{crit}$ an inflection point appears, which corresponds to a marginally stable light ring. b) For $a_2^{crit}<a_2<0$ a maximum and a minimum exist, representing a stable and an unstable light ring with respect to radial perturbations. c) For $a_2\geq0$ a minimum exists corresponding to an unstable light ring.}
\end{figure*}

\begin{figure*}[tb]
    \centering
    \setlength{\subfigheight}{2in}
    \begin{subfigure}[t]{0.4\textwidth}
        \includegraphics[width=\textwidth, height=\subfigheight]{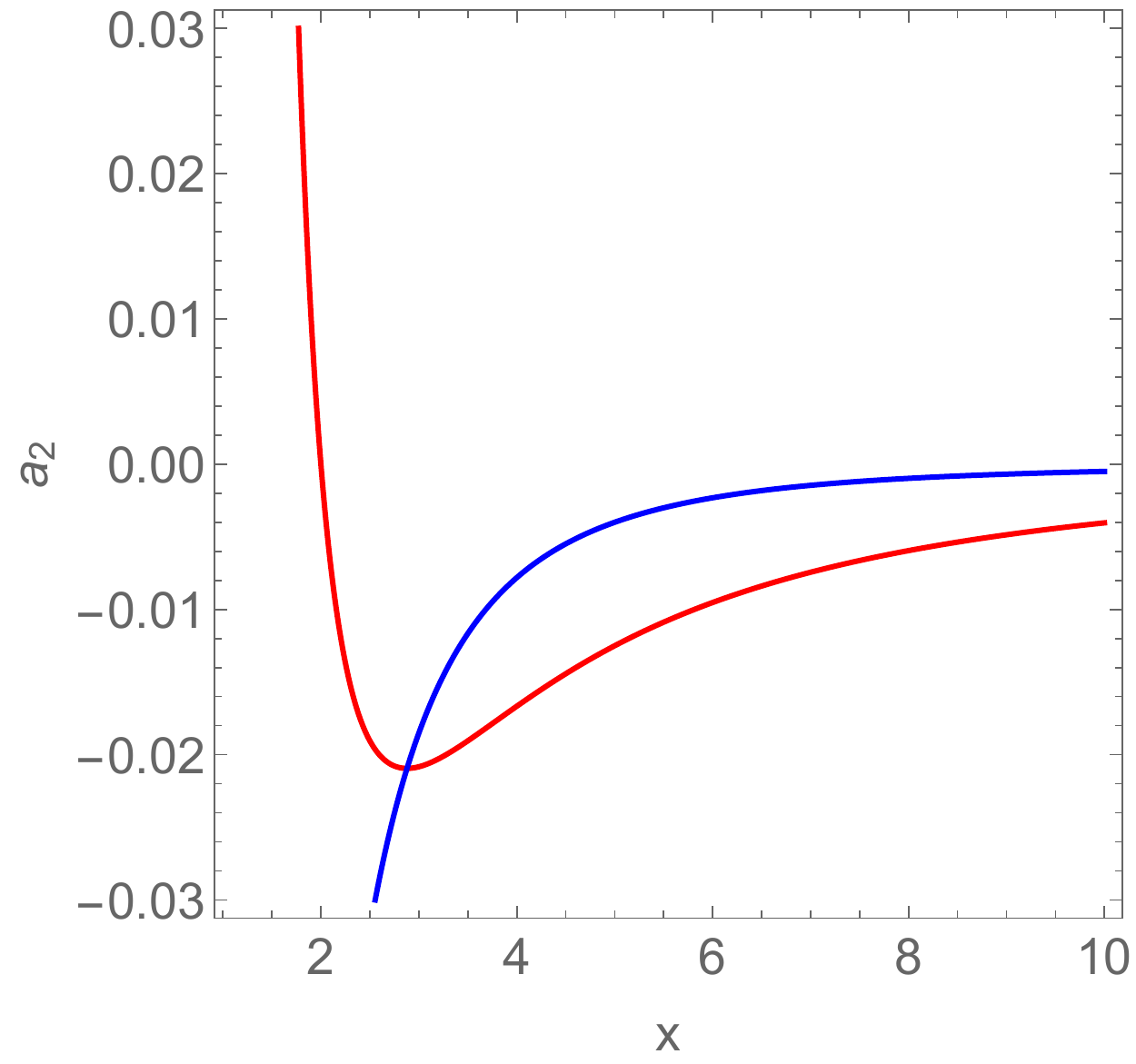}
        \caption{Light rings in the equatorial plane}\label{st}
    \end{subfigure}
    \hspace{1cm}
    \begin{subfigure}[t]{0.4\textwidth}
        \includegraphics[width=\textwidth, height=\subfigheight]{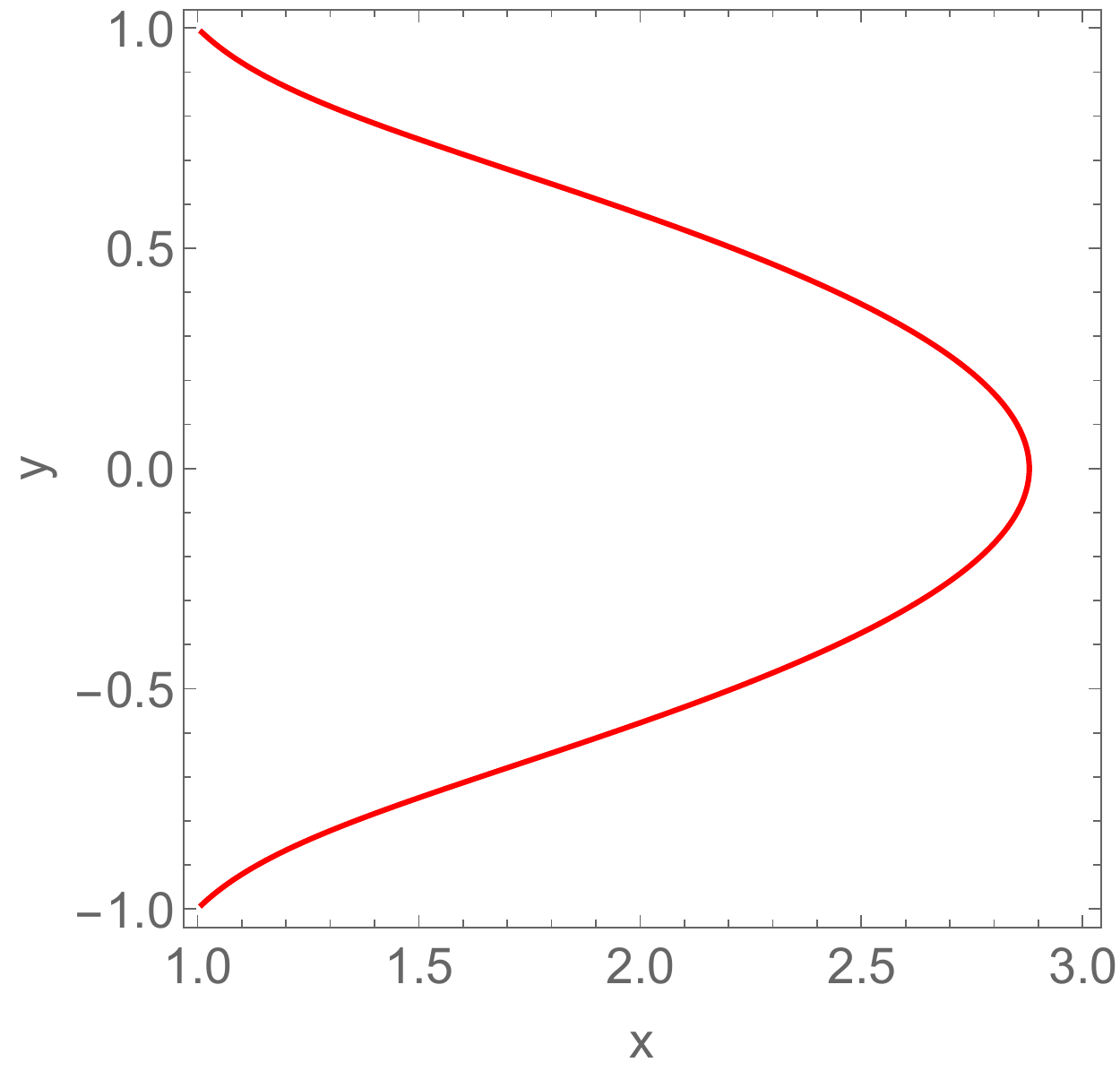}
        \caption{Light rings off the equatorial plane}\label{exist_off}
    \end{subfigure}
    \caption{Existence of circular null geodesics. Circular orbits are located on the red curves defined by ($\ref{LR0}$) and ($\ref{LRy}$), respectively, for equatorial/off-equatorial light rings. The blue curve determines the regions of stability of the light rings with respect to radial and vertical perturbations (see the Appendix). Light rings below the blue curve are stable with respect to radial perturbations, and unstable with respect to vertical ones, while in the region above the opposite is true. The two curves intersect at the location of the ISCO.}
\end{figure*}

The behavior of the potential $h^2(x,0)$ in the equatorial plane is presented in Fig. $\ref{h0}$.  Note that maxima of the potential $h^2(x,0)$ in the equatorial plane correspond to minima of the effective potential $V(x,0)$. Therefore, light rings, which are stable with respect to radial perturbations, correspond to the maxima of $h^2(x,0)$, while unstable light rings correspond to the minima of $h^2(x,0)$. Stable light rings with respect to radial perturbations exist for $x_{crit}<x< +\infty$, where $x_{crit} \approx 2.879$. The infinitely remote stable circular orbit corresponds to the limit, when the quadrupole moment vanishes, and the solution approaches the Schwarzschild black hole, while the critical value of the quadrupole moment $a_2^{crit} \approx -0.021$ determines the location of the marginally stable light ring $x_{crit} = x_{ISCO}$. Thus, adding a negative quadrupole moment with values in the  range $-0.021\leq a_2<0$ leads to moving the stable light ring from infinitely remote to the marginal location $x_{ISCO} \approx 2.879$. At the same time the coexisting unstable light ring recedes from the horizon, changing its position from $x=2$ for $a_2=0$ to $x_{ISCO}$. The domain of existence and stability of the light rings in the equatorial plane as a function of the quadrupole moment is illustrated in Fig. $\ref{st}$. We present the analysis of the light ring stability with respect to small perturbations in radial and vertical directions in the Appendix. The main result is that light rings, which are stable with respect to radial perturbations, are unstable with respect to vertical ones, and vice-versa. It is consistent with a general theorem stating that no stable light rings can exist (with respect to both radial and vertical perturbations) for vacuum solutions in general relativity \cite{Dolan:2016}.

\subsection{Light rings out of the equatorial plane}

Light rings lying outside the equatorial plane are determined by the equations $p_x = p_y =0$, $\dot{p_x}= \dot{p_y}=0$, and  $y=const. \neq 0$.
They are located at the stationary points of the 2-dimensional potential $V(x,y)$, in which it vanishes. We can equivalently obtain the stationary points of the potential $h^2(x, y)$, which corresponds to solving the equations $\partial_xh^2(x, y)=\partial_yh^2(x, y)=0$, and excluding the trivial solutions giving the symmetry axis $y=\pm 1$. Calculating the derivatives
\begin{eqnarray}
\partial_x h^2(x,y) &=& \frac{2h^2(x,y)}{x^2-1}\left[x-2-2a_2\,x(x^2-1)(3y^2-1)\right], \nonumber \\
\partial_y h^2(x,y) &=& \frac{2h^2(x,y)y}{y^2-1}\left[1+2a_2\,(1-y^2)(3x^2-1)\right],
\end{eqnarray}
we obtain that the position of the light rings outside of the equatorial plane are determined by the system
\begin{eqnarray}\label{LR}
&&x-2-2a_2\,x(x^2-1)(3y^2-1) = 0, \nonumber \\
&&1+2a_2\,(1-y^2)(3x^2-1)= 0,
\end{eqnarray}
which depends on the value of the quadrupole moment. The system can be reduced to the form

\begin{eqnarray}\label{LRx}
&&1+2a_2\,x(x^2 -y^2)=0, \nonumber \\
&&6a_2\,x^5 -8a_2\,x^3 + 3x^2 + 2a_2\,x - x -1 = 0,
\end{eqnarray}
where we should take into account that viable solutions for the light rings' positions should take the ranges $x> 1$, and $-1< y < 1$. From the first equation it is obvious that such solutions can exist only for negative values of the quadrupole moment. The second equation determines the position as a function of the quadrupole moment $x(a_2)$, which cannot be found in closed form for a general $a_2$. However, we can  conclude that a single solution exists for every $a_2$ in the range $x\in (1, +\infty)$, by examining the function $a_2(x)$. In this range the function $a_2(x)$ is monotonically increasing, and satisfies $\lim a_2(x)_{x\rightarrow1} = -\infty$, $\lim{a_2(x)}_{x\rightarrow\infty} = 0$.

Substituting the expression for $a_2(x)$ in the first equation, we obtain the relation

\begin{eqnarray}\label{LRy}
y^2 = - \frac{(x^3-3x^2+1)}{(3x^2-x-1)}\leq 1.
\end{eqnarray}
It defines a curve $y^2(x)$, which contains the positions of the light rings for any quadrupole moment (see Fig. $\ref{exist_off}$). The function $y^2(x)$ is monotonically decreasing and reaches its upper limit $y^2\rightarrow1$ when $x \rightarrow 1$, and its lower limit $y^2\rightarrow0$ when $x \rightarrow x_{ISCO} \approx 2.879$. Consequently,  we obtain limits for the possible positions of the light rings, i.e. they should satisfy $x\in (1, 2.879)$.  Taking into account that the relation $a_2(x)$ is monotonic, they put constraints on the values of $a_2$, for which solutions to the system ($\ref{LR}$) can exist. Thus, light rings lying outside the equatorial plane occur only for a quadrupole moment in the range $a_2\in (-\infty, -0.021)$. The lower limit is determined by the behavior of the function $a_2(x)$ when $x \rightarrow 1$, and the upper limit when $x \rightarrow x_{ISCO}$.

In summary, every value of $a_2$ in the range $a_2\in (-\infty, -0.021)$ determines a pair of light rings located symmetrically with respect to the equatorial plane. Their coordinates are given by $(x, \pm y)$, where $x$ is a solution to ($\ref{LRx}$),  and $y$ is obtained by ($\ref{LRy}$). Increasing the quadrupole moment, the value of $x$ increases, while the value of $|y|$ decreases. Thus, the light ring pairs move away from the horizon and simultaneously approach the equatorial plane. In the limit $x\rightarrow 2.879$,  $y^2\rightarrow0$,  they merge into the marginally stable light ring in the equatorial plane. We should also note that the regions of existence of the light rings in and outside the equatorial plane are complimentary, as light rings in the equatorial plane exist only for quadrupole moments in the range $a_2\in ( -0.021, +\infty)$. Referring to the general result for stationary and axisymmetric vacuum spacetimes \cite{Dolan:2016}, all the light rings should be unstable, and we observe that they correspond to saddle points of the effective potential $h^2(x,y)$.

\begin{figure*}[tb]
    \centering  
    \setlength{\subfigheight}{3in}
    \begin{subfigure}[t]{0.49\textwidth}
        \includegraphics[width=\textwidth, height=\subfigheight]{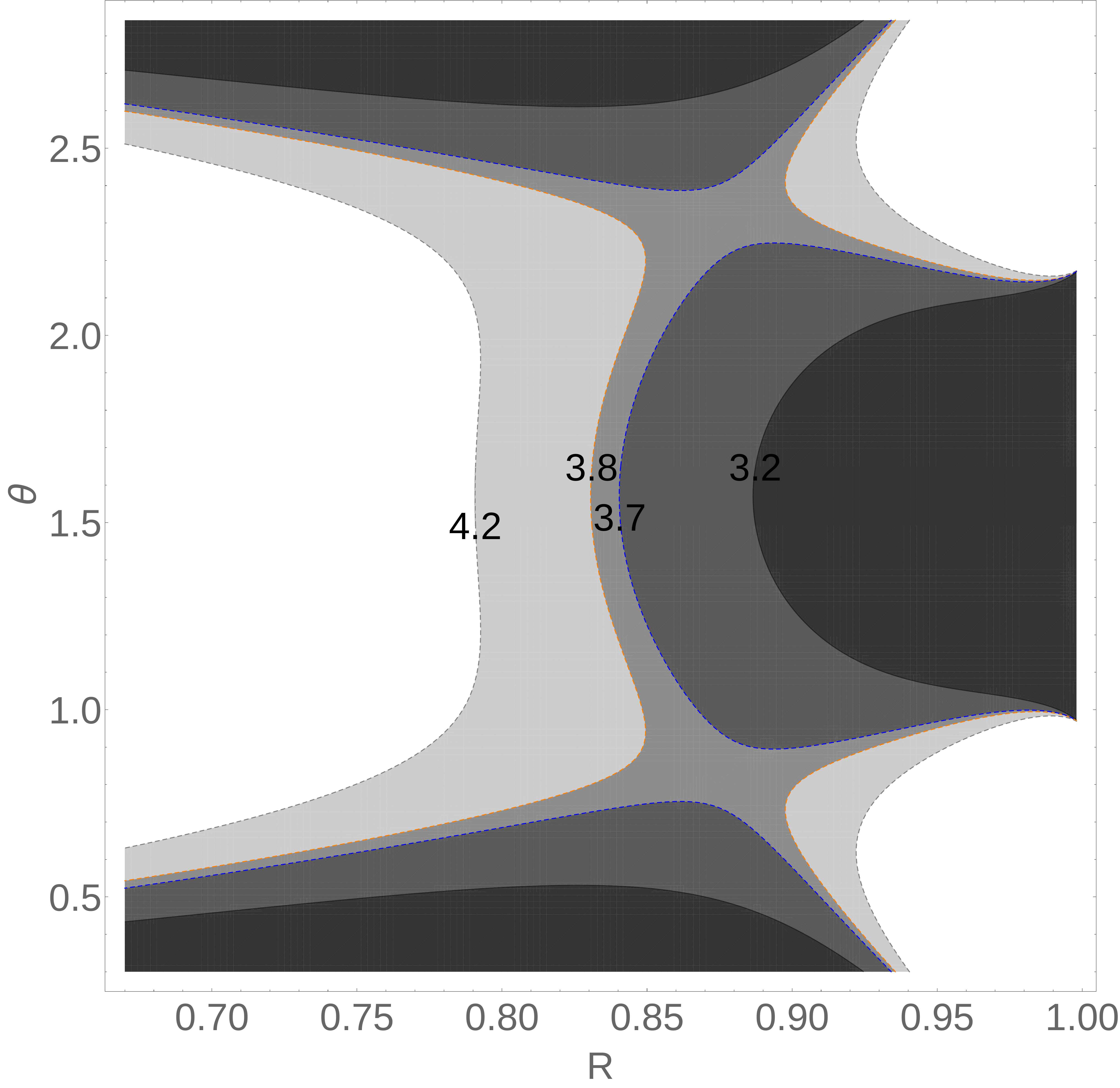}
        \caption{$a_2= -0.12$}\label{Contour1a}
    \end{subfigure}
    %\hspace{1cm}
    \begin{subfigure}[t]{0.49\textwidth}
        \includegraphics[width=\textwidth, height=\subfigheight]{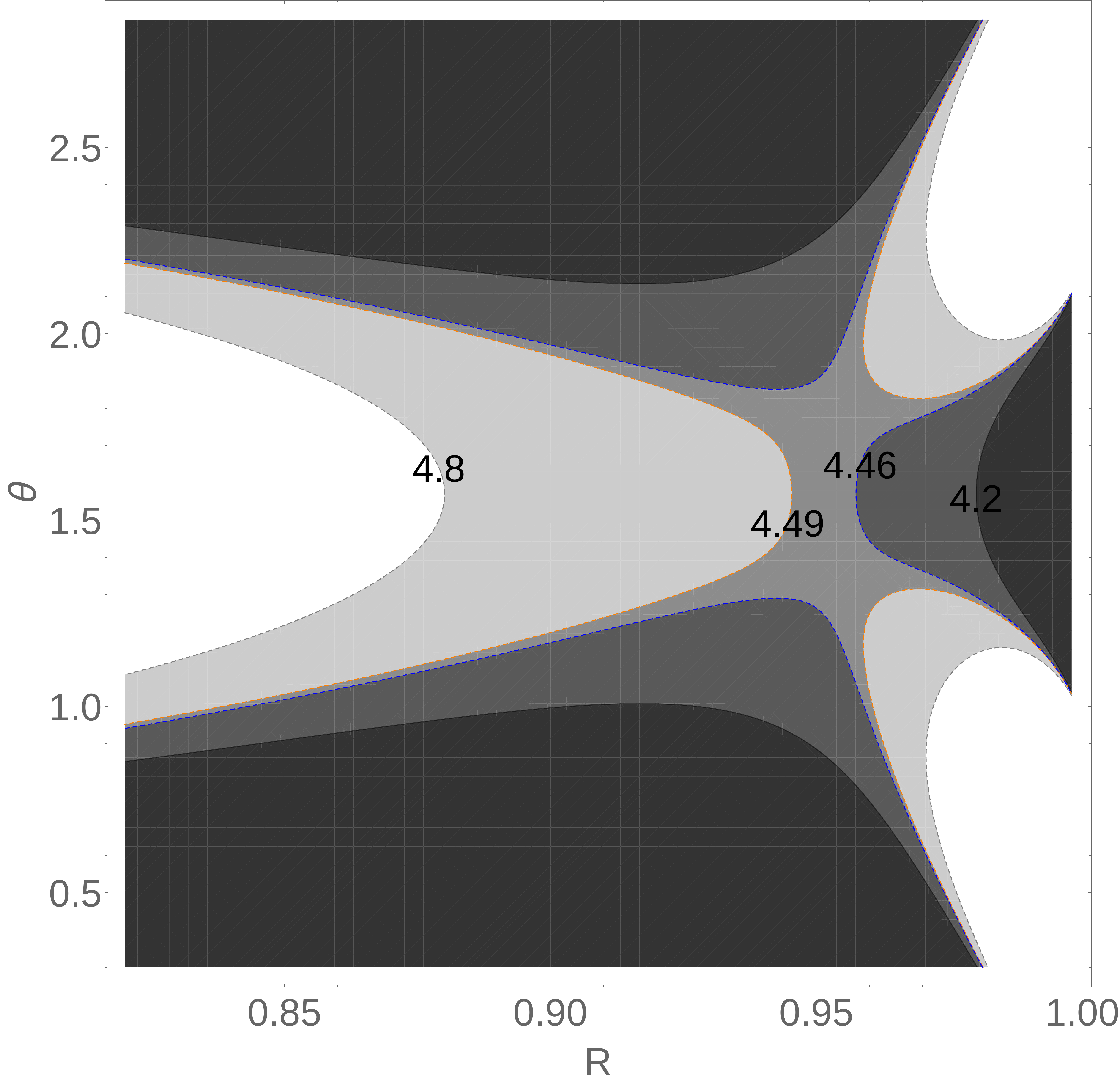}
        \caption{$a_2= -0.024$}\label{Contour1b}
    \end{subfigure}
 \newline \hspace{0.5cm}
    \begin{subfigure}[t]{0.49\textwidth}
        \includegraphics[width=\textwidth, height=\subfigheight]{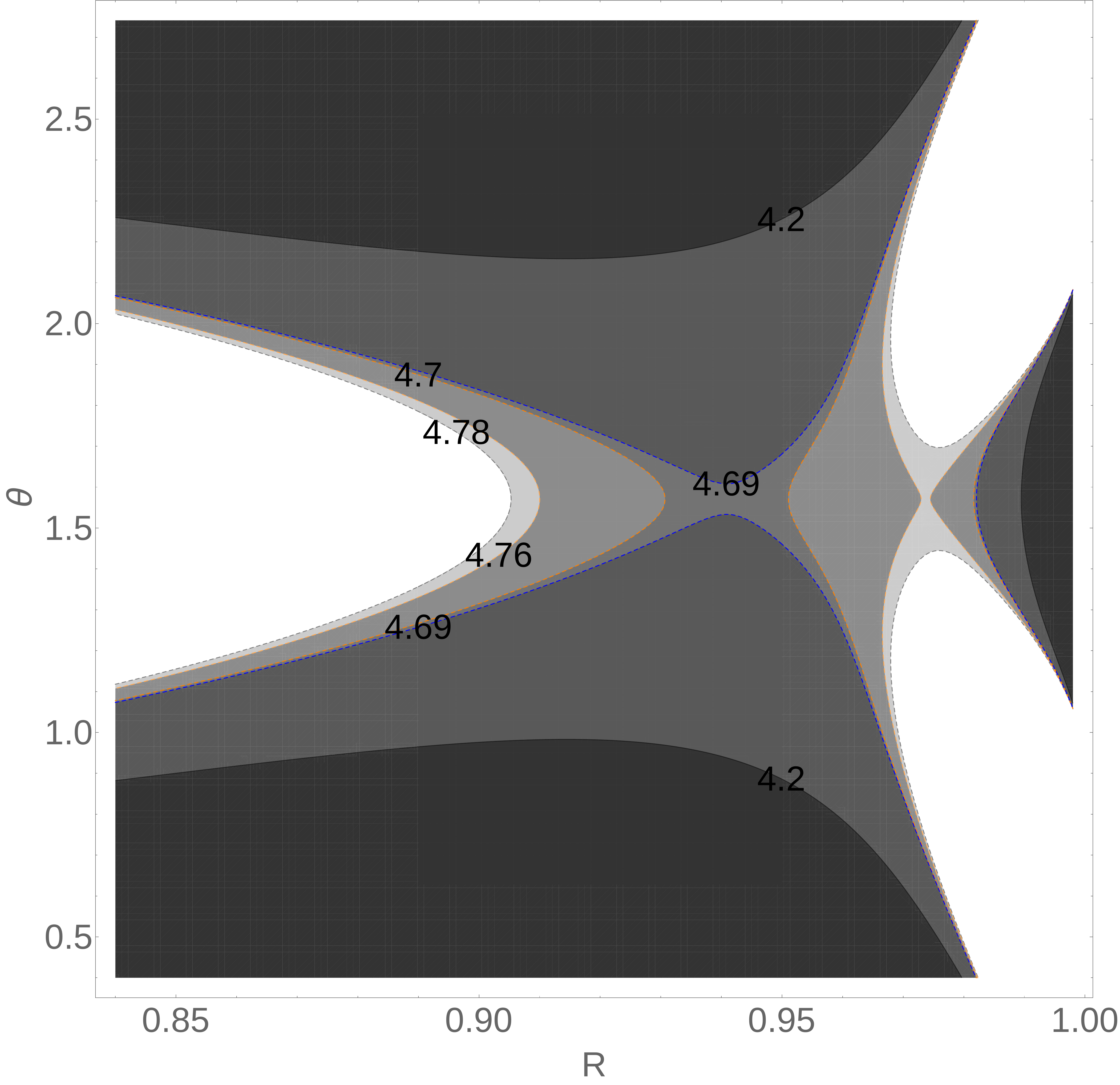}
        \caption{$a_2=-0.018$}\label{Contour1c}
    \end{subfigure}
    %\hspace{1cm}
    \begin{subfigure}[t]{0.49\textwidth}
       \includegraphics[width=\textwidth, height=\subfigheight]{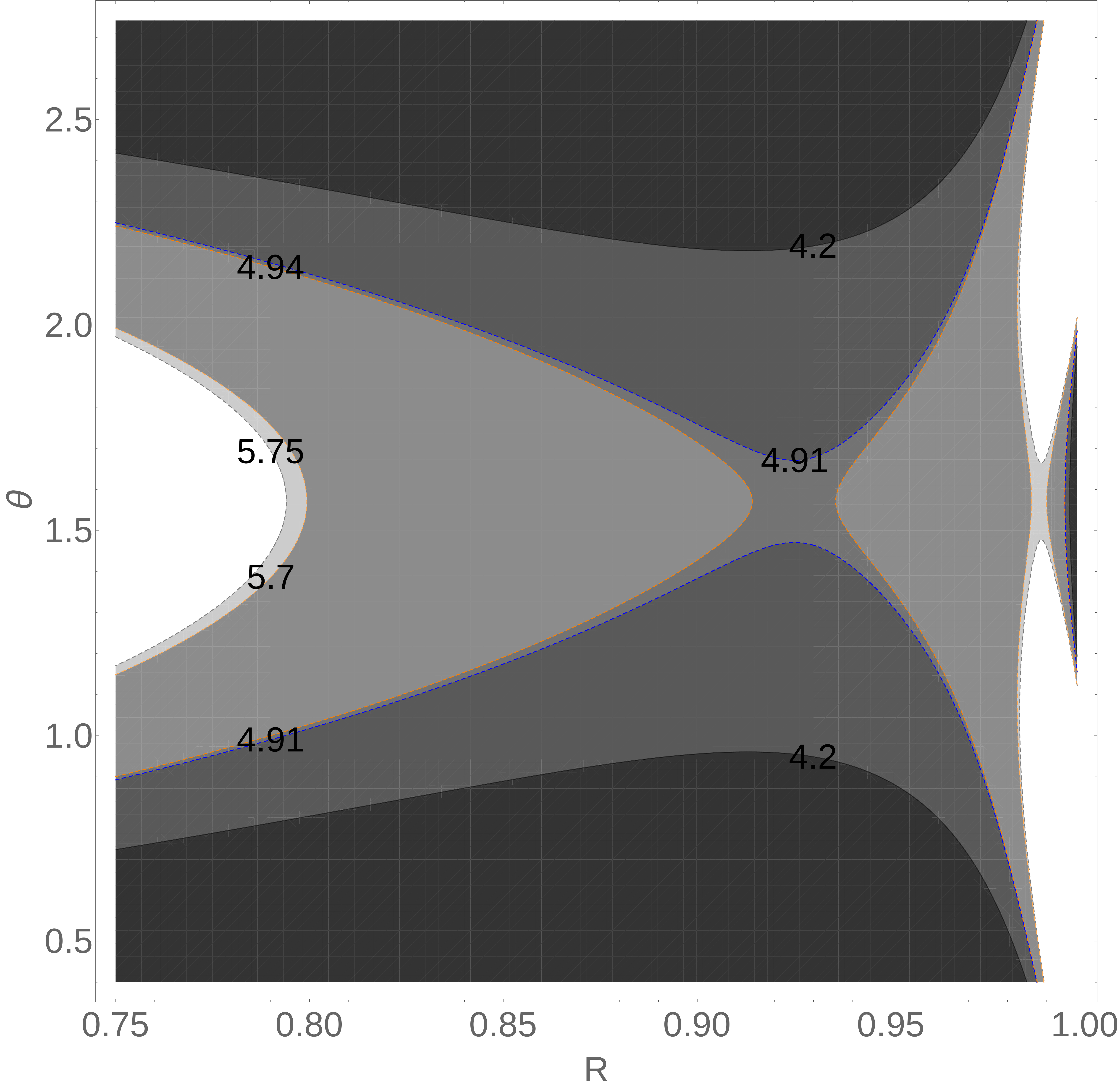}
        \caption{$a_2= -0.01$}\label{Contour1d}
    \end{subfigure}
    \caption{\label{Contour1}Contour plots of the potential $h(x,y)$ for negative quadrupole moments.  In the range $a_2\in (-\infty, -0.021)$ two saddle points are observed, located symmetrically with respect to the equatorial plane ((a) and (b)). Decreasing the quadrupole moment, the light rings move away from the equatorial plane, and approach the horizon. In the range $a_2\in ( -0.021, 0)$ the saddle points lie in the equatorial plane ((c) and (b)). Decreasing  $a_2$, the outer light ring moves closer to the horizon, and approaches the inner one.}
\end{figure*}

\begin{figure*}[tb]
    \centering
    \setlength{\subfigheight}{3in}
    \begin{subfigure}[t]{0.49\textwidth}
        \includegraphics[width=\textwidth, height=\subfigheight]{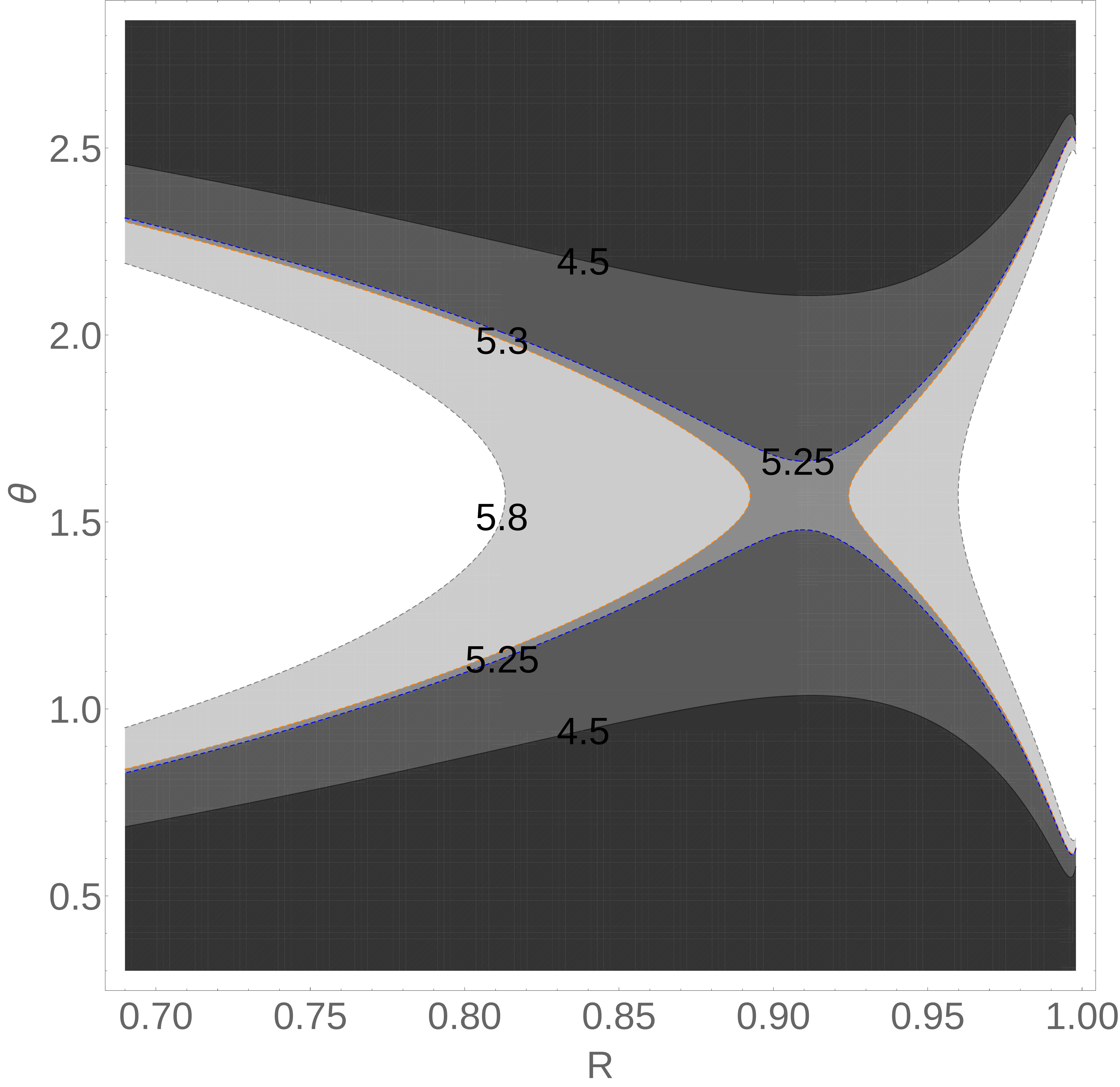}
        \caption{$a_2=0.003$}
    \end{subfigure}
    %\hspace{1cm}
    \begin{subfigure}[t]{0.49\textwidth}
        \includegraphics[width=\textwidth, height=\subfigheight]{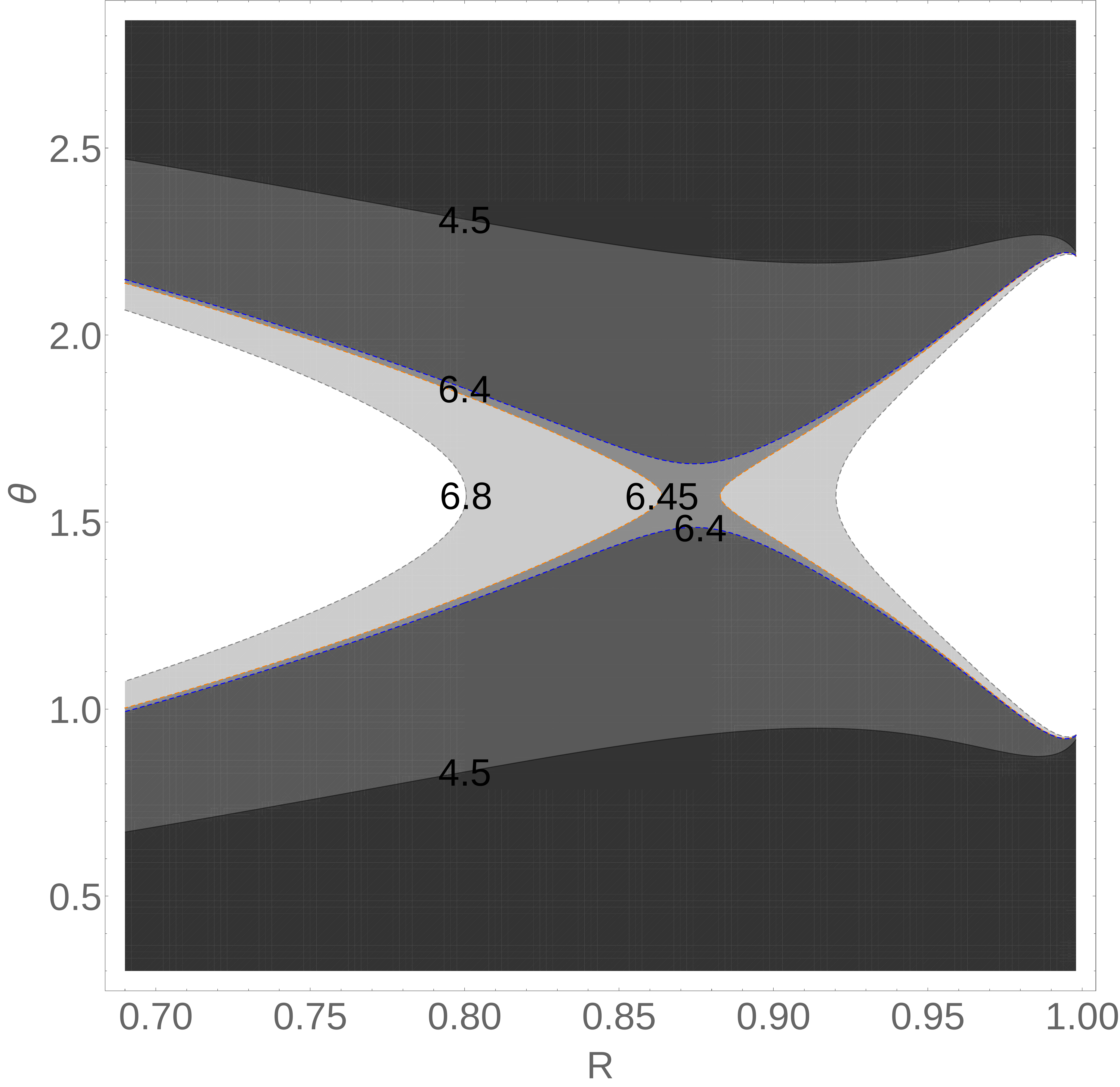}
        \caption{$a_2= 0.05$}
    \end{subfigure}
    \caption{\label{Contour3}Contour plots of the potential $h(x,y)$ for positive quadrupole moments. A single saddle point is observed lying in the equatorial plane. Increasing the quadrupole moment $a_2$, the light ring approaches the horizon.}
\end{figure*}

In Figs. $\ref{Contour1}$ and $\ref{Contour3}$ we present contour plots of the potential $h(x,y)$ for the three sets of ranges of the quadrupole moment, which lead to qualitatively different light ring structures. In each set we illustrate the transformation of the potential when varying the quadrupole moment. We introduce a compactified radial coordinate defined by means of the horizon position $x_H$

\begin{equation}
{\rm R} = \frac{\sqrt{r^2-r_H^2}}{1+ \sqrt{r^2-r_H^2}} =  \frac{\sqrt{(x+1)^2-(x_H+1)^2}}{1+ \sqrt{(x+1)^2-(x_H+1)^2}},
\end{equation}
which maps the domain of outer communication $x\in[1, +\infty)$ to the finite interval ${\rm R}\in[0,1)$. For giving a better intuition we use the polar angle $\theta$ instead of the spheroidal coordinate $y=\cos\theta$.

In Figs. $\ref{Contour1a}$ and $\ref{Contour1b}$ the quadrupole moment belongs to the range $a_2\in (-\infty, -0.021)$.  Two saddle points are observed, corresponding to light rings located symmetrically with respect to the equatorial plane $y=0$. We see that when we decrease the quadrupole moment $a_2$, the light rings move away from the equatorial plane, and approach the horizon.

In both plots four representative contours of the potential $h(x,y)$ are illustrated in colour. Each of them represents the boundary of the forbidden region for null rays with impact parameter equal to the value of the potential at the boundary contour, i.e. $\eta = h_{bound}$. The particular value of $\eta$ is denoted on the relevant contour. Each contour separates regions filled in darker hues of gray from regions filled in lighter shades. The convention is that for each $\eta$ the forbidden regions correspond to the relatively darker areas.

In Figs. $\ref{Contour1c}$ and $\ref{Contour1d}$ the quadrupole moment belongs to the range $a_2\in (-0.021, 0)$, and the saddle points are located only in the equatorial plane as expected.  When we decrease the quadrupole moment $a_2$, we see that the outer light ring moves closer to the horizon, while the inner one deviates from it, so both approach each other. Certain representative contours are depicted in colour. We follow the same convention as in the previous case in Fig. $\ref{Contour1a}$ and  $\ref{Contour1b}$. Each  contour is the boundary of the forbidden region for light rays with impact parameter $\eta = h_{bound}$, and forbidden regions correspond to relatively darker areas.

In Fig. $\ref{Contour3}$ we illustrate the effective potential for positive quadrupole moments $a_2\in (0, +\infty)$. The left panel corresponds to small positive values, while the right one represents the evolution of the potential when the quadrupole moment is increased. A single light ring in the equatorial plane is observed, which moves closer to the horizon for higher quadrupole moments.

In the following analysis we will not consider the case of positive quadrupole moments, as, according to ($\ref{SEC}$), it corresponds to interaction with external material sources that violate the strong energy condition. We will concentrate on the more realistic case of negative quadrupole moments.

\subsection{\label{sub:potentials}Effective potentials for $a_2 <0$}

\begin{figure*}[tb]
    \centering
    \begin{subfigure}[t]{0.45\textwidth}
        \centering
        \includegraphics[width=0.95\textwidth]{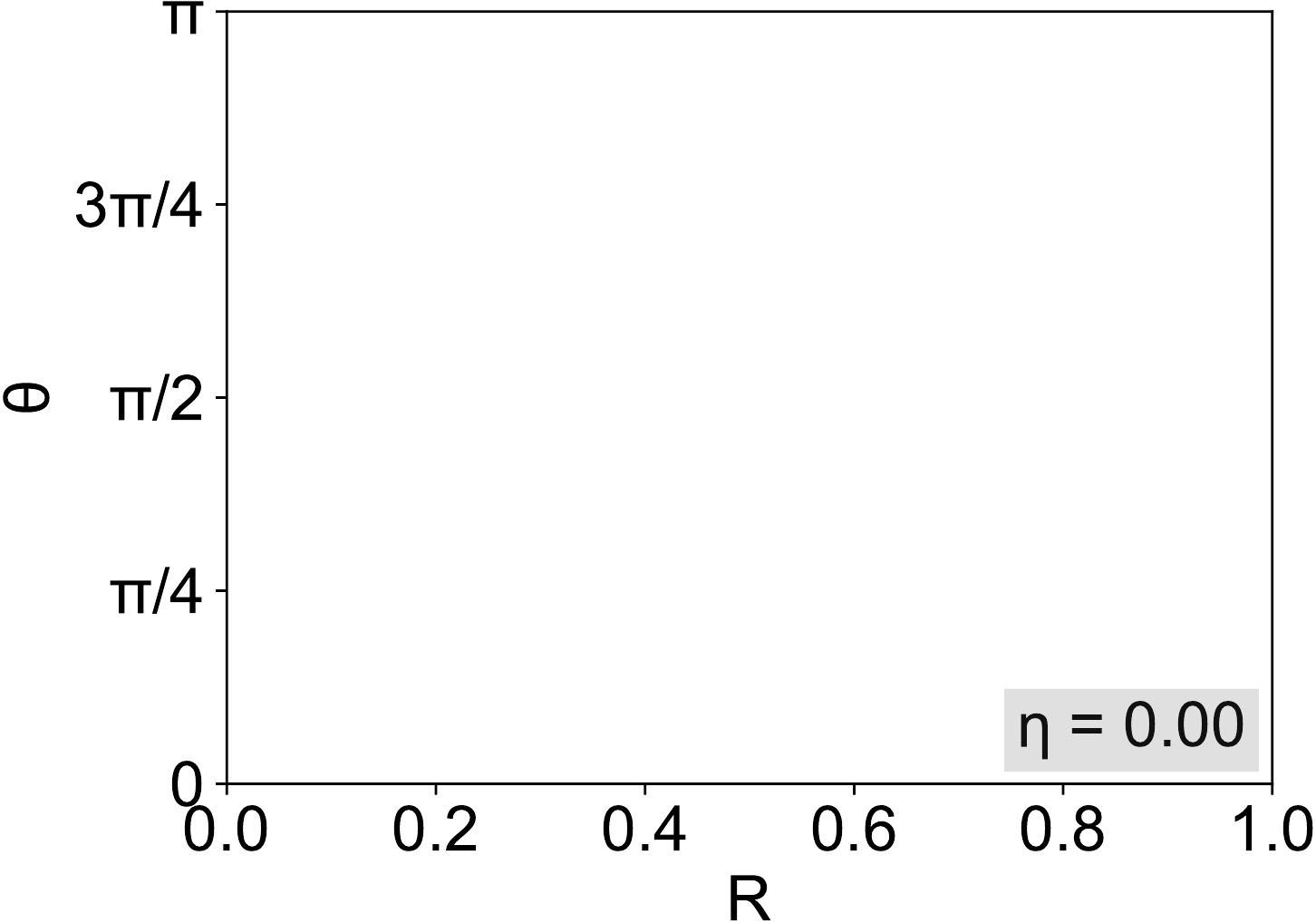}
        \caption{$\eta=0$}
    \end{subfigure}
    %\\[5mm]
    \begin{subfigure}[t]{0.45\textwidth}
        \centering
        \includegraphics[width=0.95\textwidth]{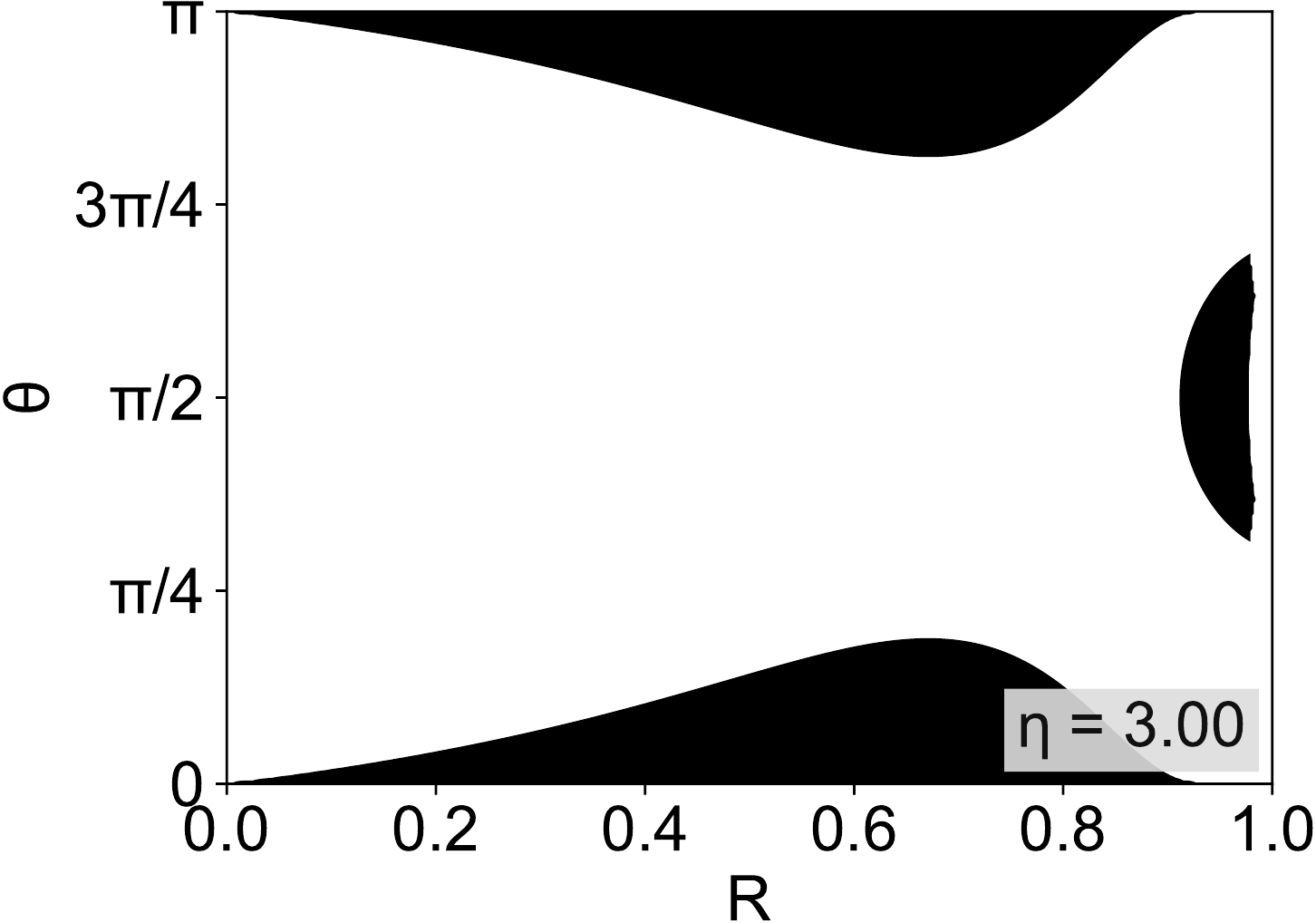}
        \caption{$0<\eta<\eta_1$}
    \end{subfigure}
    \\%[5mm]
    \begin{subfigure}[t]{0.45\textwidth}
        \centering
        \includegraphics[width=0.95\textwidth]{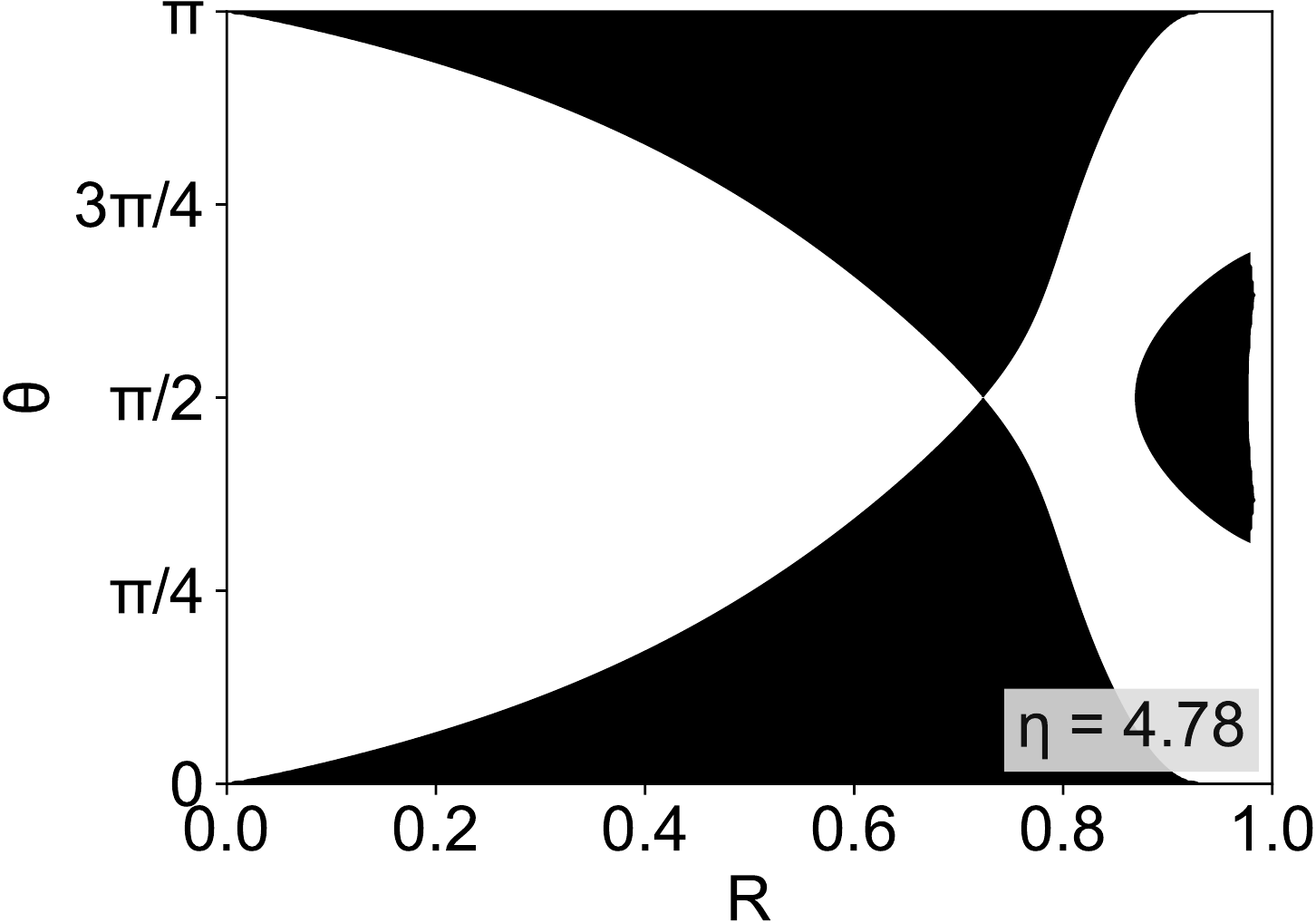}
        \caption{$\eta=\eta_1$ (radially unstable light ring)}
    \end{subfigure}
    %\\[5mm]
    \begin{subfigure}[t]{0.45\textwidth}
        \centering
        \includegraphics[width=0.95\textwidth]{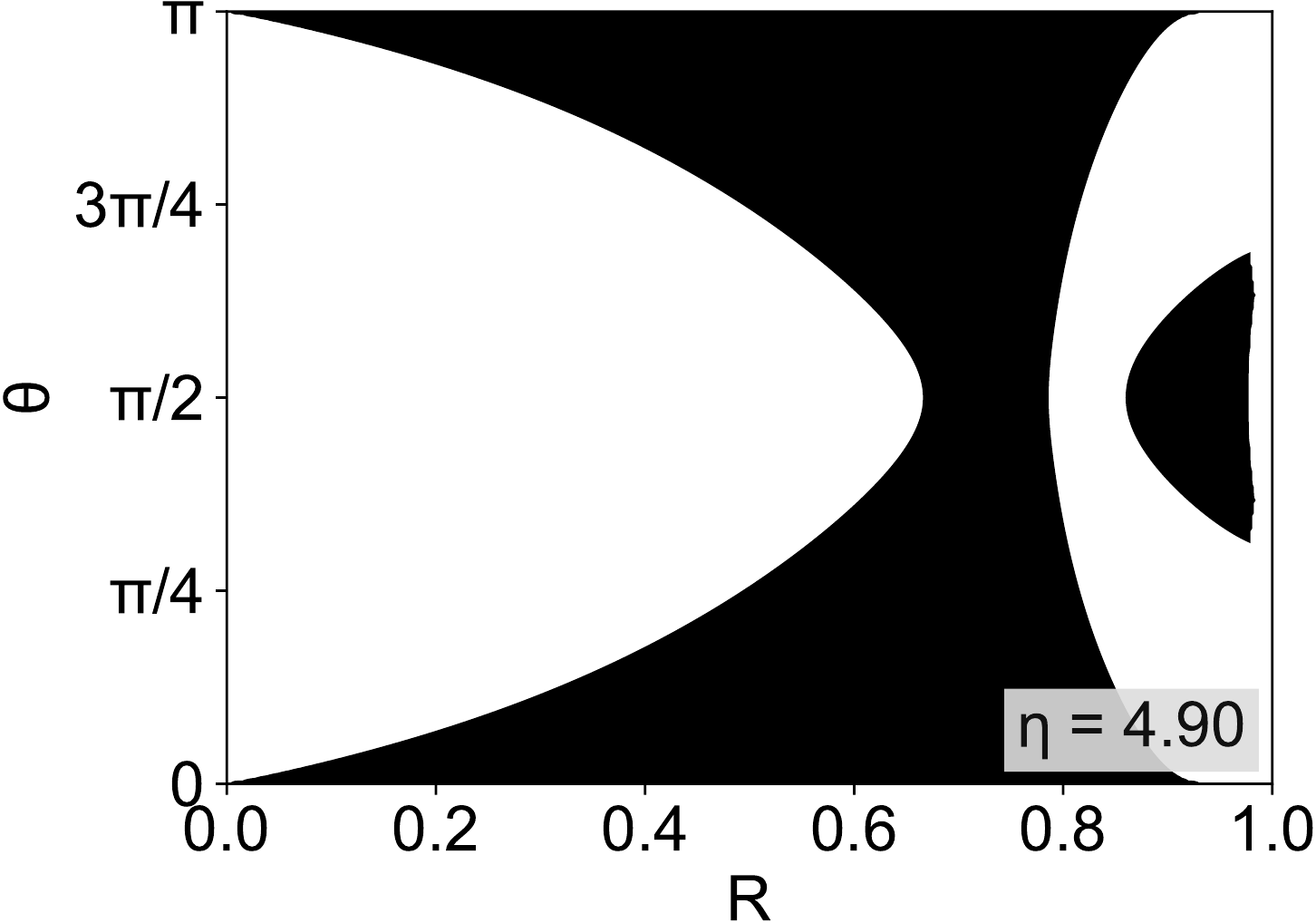}
        \caption{$\eta_1<\eta<\eta_2$}
    \end{subfigure}
    \\%[5mm]
    \begin{subfigure}[t]{0.45\textwidth}
        \centering
        \includegraphics[width=0.95\textwidth]{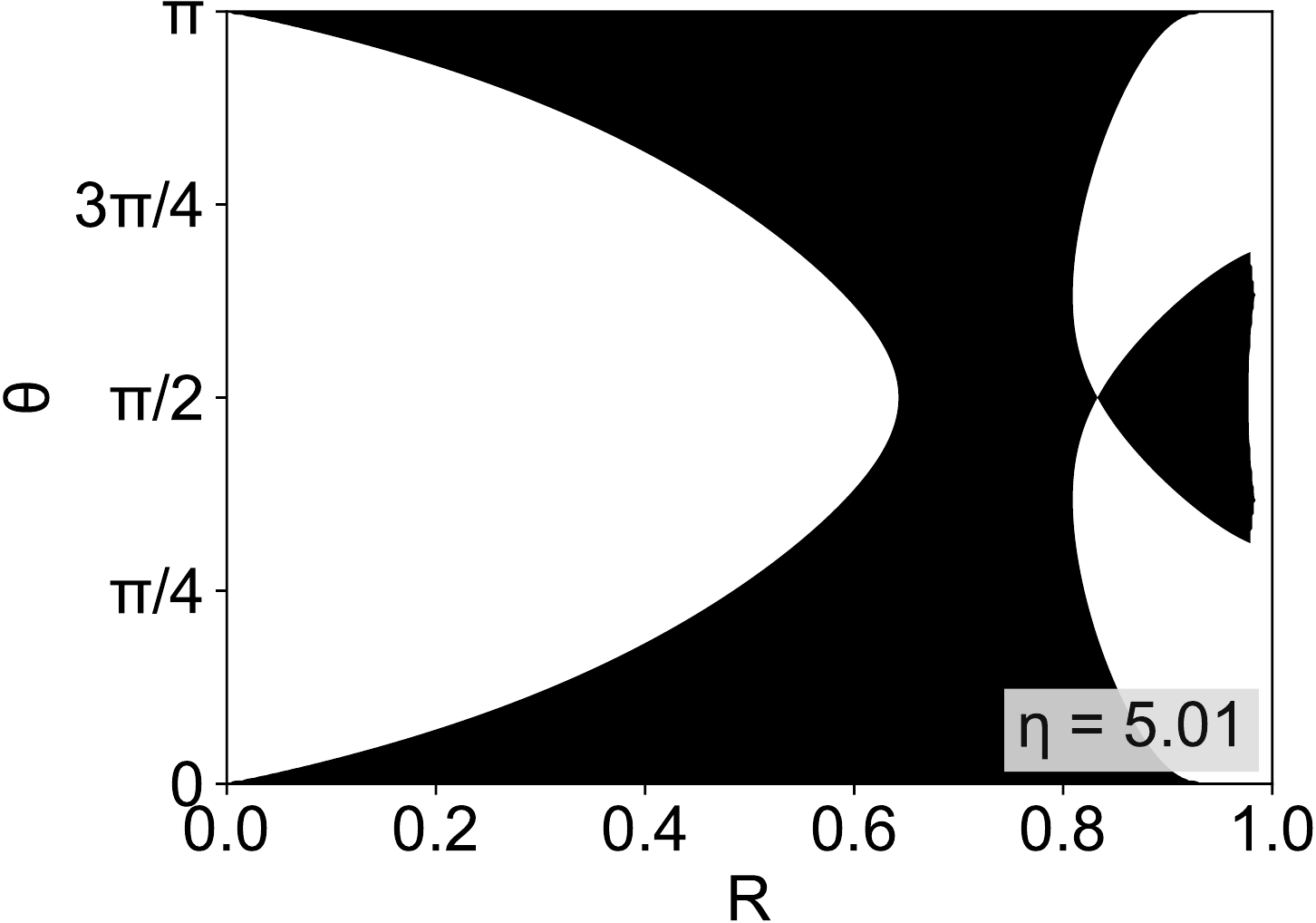}
        \caption{$\eta=\eta_2$ (radially stable light ring)}
    \end{subfigure}
    %\\[5mm]
    \begin{subfigure}[t]{0.45\textwidth}
        \centering
        \includegraphics[width=0.95\textwidth]{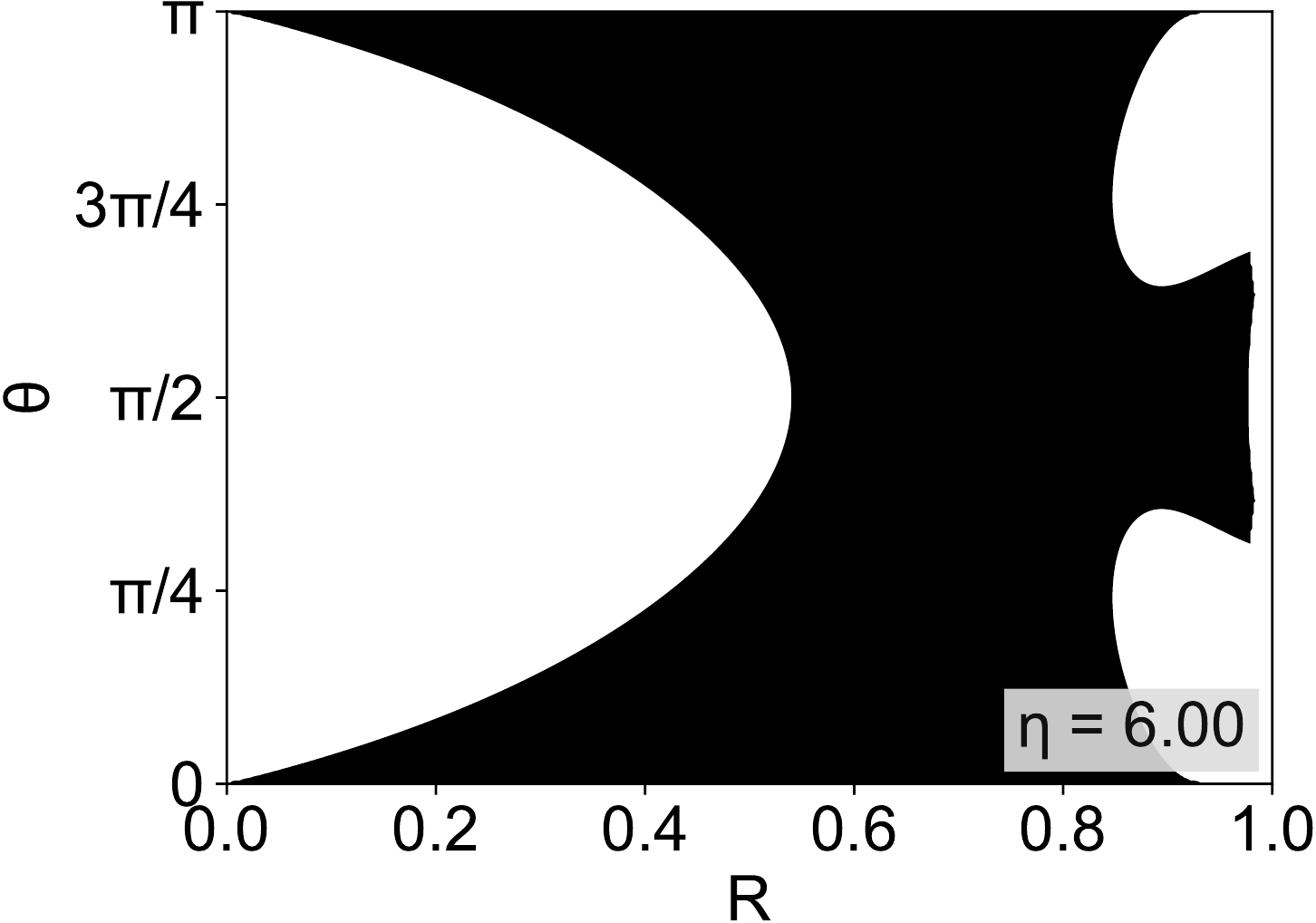}
        \caption{$\eta_2<\eta$}
    \end{subfigure}

    \caption{\label{fig:potentials1}Effective potentials for quadrupole moment $a_2=-0.015>a_2^{crit}$ and various values of impact parameter $\eta$ illustrating the effect of increasing $|\eta|$.}
\end{figure*}

\begin{figure*}[tb]
    \centering
    \begin{subfigure}[t]{0.45\textwidth}
        \centering
        \includegraphics[width=0.95\textwidth]{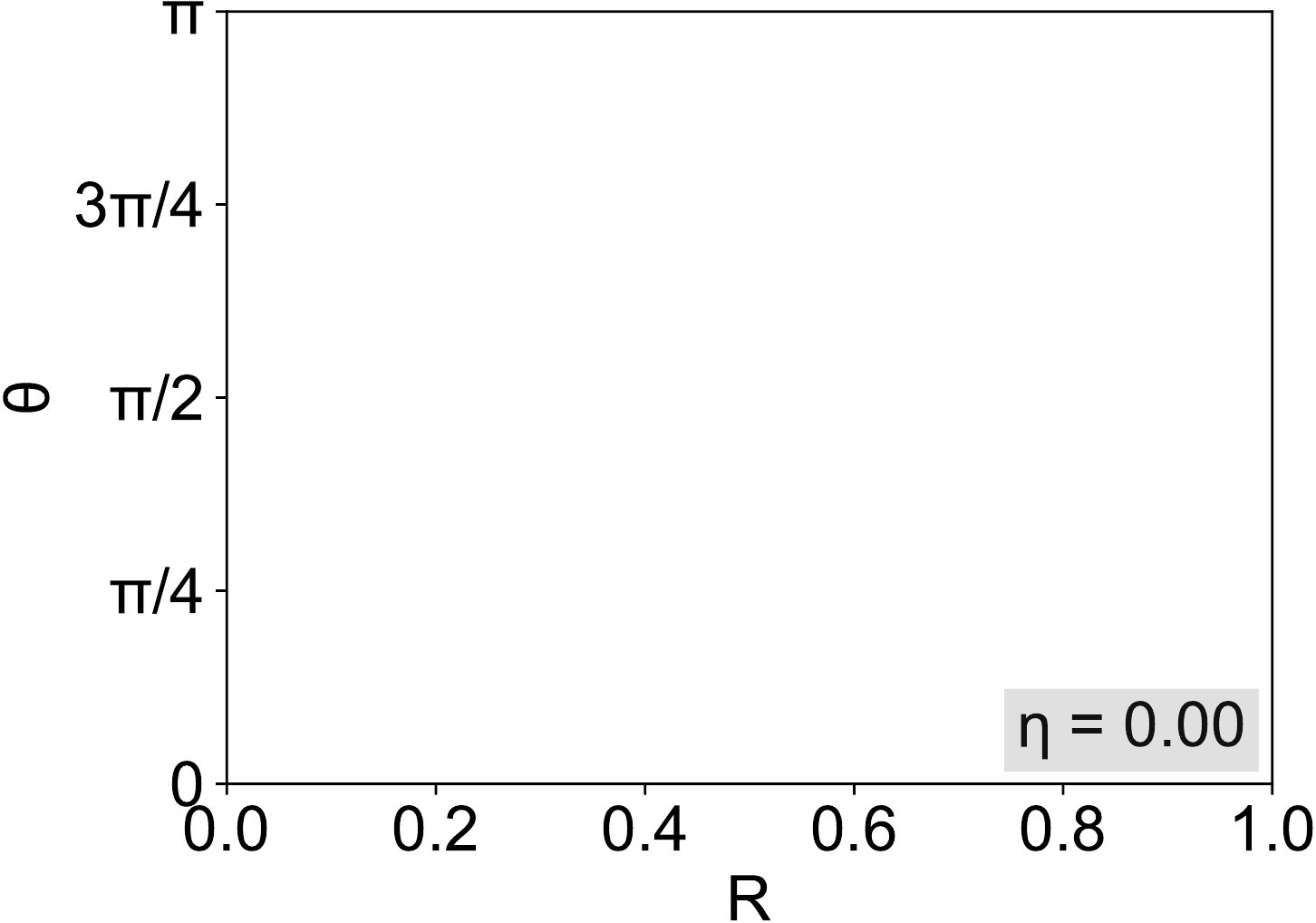}
        \caption{$\eta=0$}
    \end{subfigure}
    %\\[5mm]
    \begin{subfigure}[t]{0.45\textwidth}
        \centering
        \includegraphics[width=0.95\textwidth]{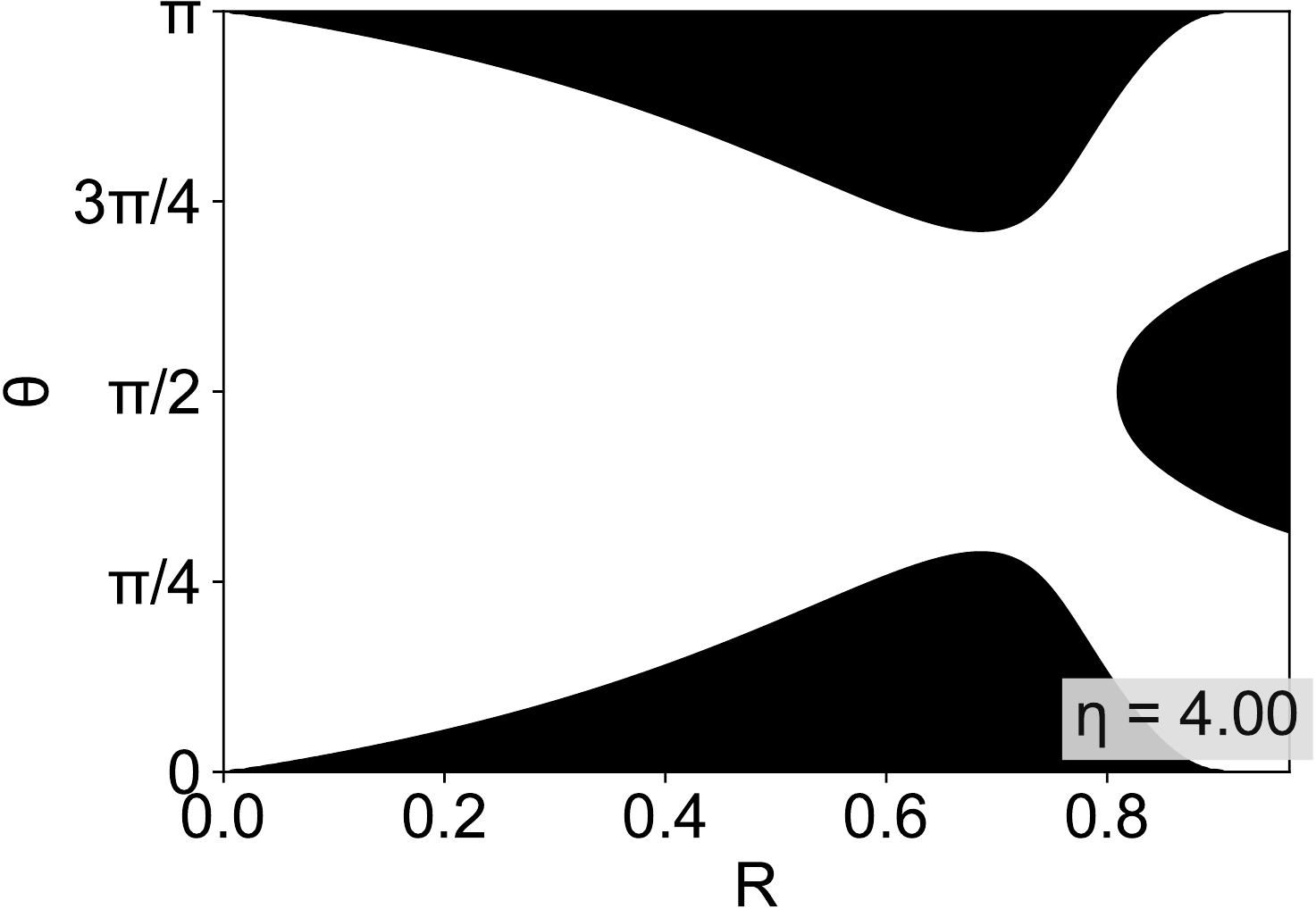}
        \caption{$0<\eta<\eta_1$}
    \end{subfigure}
    \\%[5mm]
    \centering
    \begin{subfigure}[t]{0.45\textwidth}
        \centering
        \includegraphics[width=0.95\textwidth]{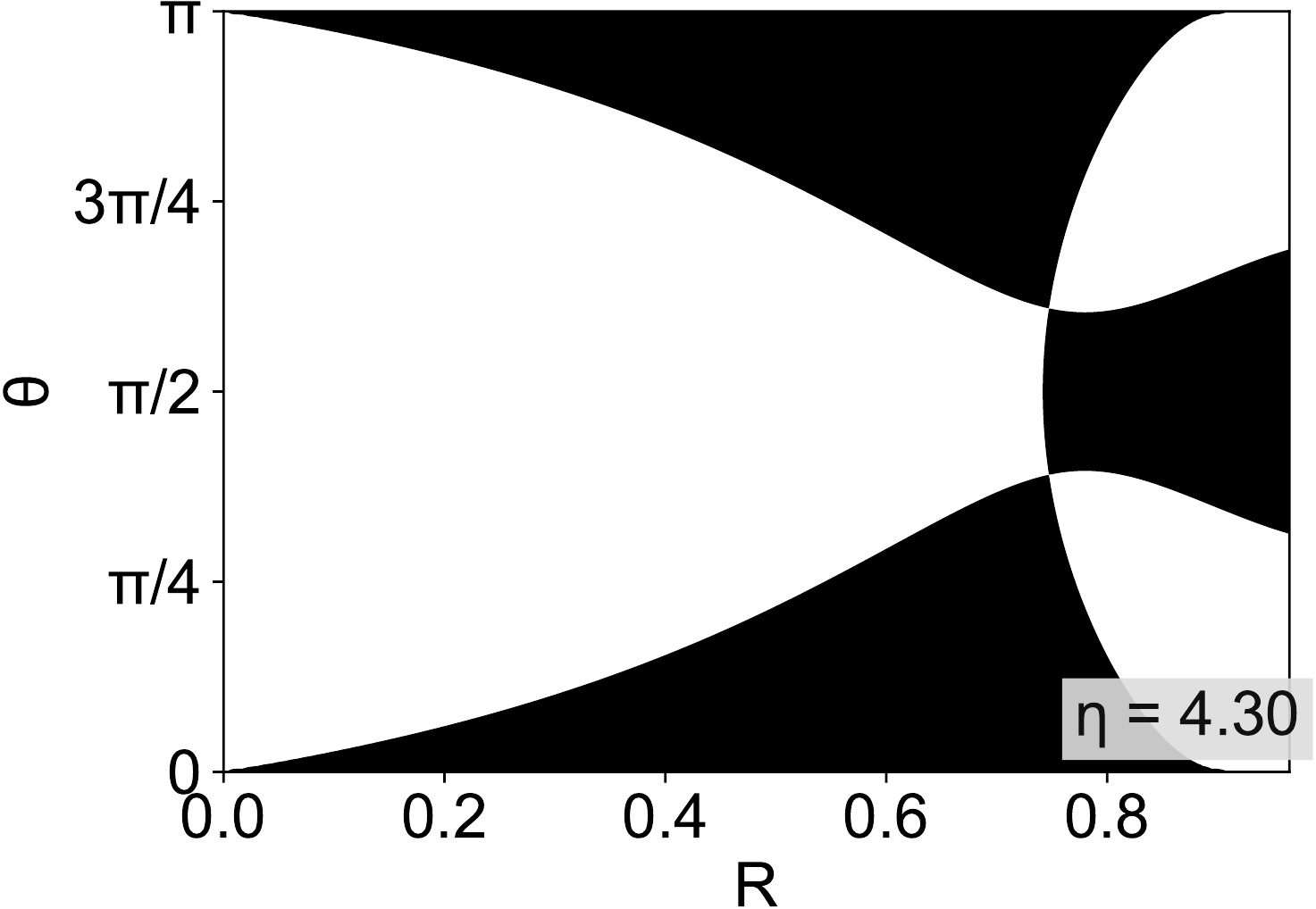}
        \caption{$\eta=\eta_1$}
    \end{subfigure}
    %\\[5mm]
    \begin{subfigure}[t]{0.45\textwidth}
        \centering
        \includegraphics[width=0.95\textwidth]{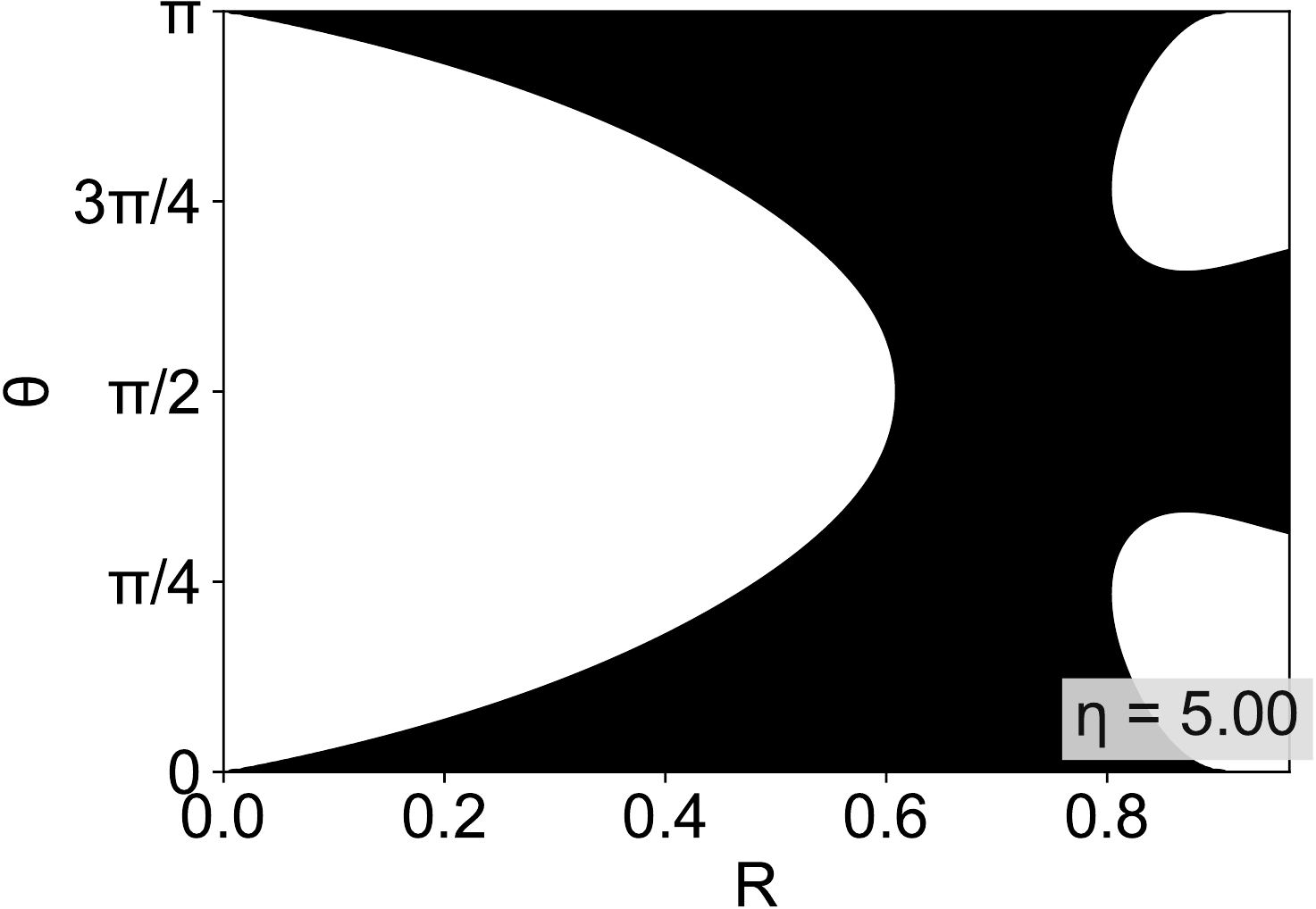}
        \caption{$\eta_1<\eta$}
    \end{subfigure}

    \caption{\label{fig:potentials2}Effective potentials for quadrupole moment $a_2=-0.03<a_2^{crit}$ and various values of impact parameter $\eta$ illustrating the effect of increasing $|\eta|$.}
\end{figure*}

The effect of the different values of the impact parameter $\eta$ on the motion is best analyzed in terms of the allowed region defined by the effective potential. Figures \ref{fig:potentials1} and \ref{fig:potentials2} show a sequence of contour plots of the effective potential for increasing values of $\eta$ with $a_2^{crit}<a_2<0$ and $a_2<a_2^{crit}<0$, respectively; in these plots dark regions are forbidden and white allowed. Note that due to the symmetry of the problem, the plots look exactly the same for $\eta$ and $-\eta$, hence plots for negative values are omitted.

As discussed in Section III, for large values of $x$ the effective potential is always positive in the vicinity of the equatorial plane, with the exception of photons with an impact parameter $\eta=0$. This leads to the formation of an equatorial forbidden region, which is absent for asymptotically flat solutions. We therefore interpret this forbidden region as arising due to the influence of the external gravitational source. The interaction with the external gravitational field is then most straightforwardly encoded in the photon dynamics through scattering from the vicinity of the equatorial forbidden region. Such trajectories are illustrated for example in Fig. $\ref{fig:trajectories1}$. In our discussion we investigate this type of interaction.

In Fig. \ref{fig:potentials1} we illustrate the behavior of the effective potential when varying the impact parameter $\eta$ for quadrupole moments in the range $a_2^{crit}<a_2<0$. We see that when  increasing $\eta$, the throat at the inner, radially unstable, light ring starts to close, while the external region begins to move inwards (b). At $\eta=\eta_1$, the throat closes at the inner light ring position $x_1$ (c). From now on, it is not possible for light rays from outside $x_1$ to reach the black hole. Increasing $\eta$ further increases the width of the barrier at $x_1$, while bringing in the external region even more (d). At $\eta=\eta_2$, the external region touches the barrier at $x_2$, the position of the outer, radially stable, light ring (e). As evident from the effective potential plot this light ring is only stable in the radial direction, while it is unstable in the $\vartheta$ direction. Increasing $\eta$ even further, the barrier and external region merge into one connected region, segmenting the allowable regions into three distinct parts (f).

For $a_2<a_2^{crit}<0$, Fig. \ref{fig:potentials2} shows an even simpler dynamic.  As $\eta$ increases the throat begins to form, and the external region moves inward (b). However, in this case the external region moves fast enough to touch the forming throat at $x_1$ for $\eta=\eta_1$ before it has a chance to fully close (c); note that at $x_1$ there are now two off-equatorial, radially unstable, light rings. In a slight abuse of terminology we shall still refer to these as inner light rings. Increasing $\eta$ further beyond this value leads to the same segmentation of the allowable region into three distinct regions also observed in the previous case (d).

For any fixed $a_2$ we can find the value of the impact parameter $\eta_1$, at which the throat associated with the inner light ring(s) closes. It coincides with the value of the potential $h(x,y)$ at the light ring position, i.e. we have $\pm\eta_1 = h_{LR}$. For $a_2>a^{crit}_2$ the inner light ring lies in the equatorial plane. Using ($\ref{LR0}$), which defines its position for a given $a_2$, we can derive an expression for the value of the potential $h_{LR}$ at the inner light ring as a function of the light ring position $x_1$

\begin{eqnarray}
h^2_{LR} = \frac{(x_1 +1)^3}{(x_1-1)}\exp\left[-\frac{(x_1-2)(x_1^2+1)}{x_1(x_1^2-1)}\right],
\end{eqnarray}
or implicitly, by using the relation $x_1=x_1(a_2)$, as a function of the quadrupole moment. For $a^{crit}_2<a_2<0$, ($\ref{LR0}$) possesses two real roots in the range $x\in(1, \infty)$, and $x_1$ is associated with the lower one. 

A similar expression can be derived for  $a_2<a^{crit}_2$, when the inner light rings arise in a symmetric pair with respect to the equatorial plane. Using ($\ref{LRx}$)-($\ref{LRy}$) we obtain for the potential $h_{LR}$ at the light rings

\begin{eqnarray}
h^2_{LR} = \frac{x_1(x_1 +1)^4}{3x_1^2-x_1-1}\exp\left[\frac{3x_1^3-6x_1^2+x_1+2}{x_1(1-3x_1^2)}\right],
\end{eqnarray}
where $x_1$ is again the inner light rings' radial coordinate, which depends implicitly on the quadrupole moment by ($\ref{LRx}$). In all the cases the throat of the effective potential associated with the inner light ring is open for impact parameters satisfying $-\eta_1<\eta<\eta_1$.

%%%%%%%%%%%%%%%%%%%%%%%%%%%%%%%%%%%%%%%%%%%%%%%%%%%%%%%%%%%%%%%%%%%%%%%%%%%%

\section{\label{sec:multipleshadows} Multiple shadows}
For negative values of the quadrupole moment $a_2<0$ we can define a sufficient condition for multiple shadows to appear. This is based on the observation that multiple shadows arise principally from processes where light rays scatter off the external forbidden region before plunging behind the horizon; while it is not clear that this is the only mechanism for multiple shadows, it is a sufficient one. When the external forbidden region is cut off, by being placed outside the celestial sphere, this dynamic is not possible. Similarly, if the throat, which is forming at the position of the unstable photon orbit and leading to the black hole is closed then shadows are not possible at all. These conditions together constrain when multiple shadows can be seen by an observer.

Let us fix the space-time under consideration by picking a value for $a_2<0$. Then, on this background, we shall assume that our observer is situated at some position $(x_o,y_o=0)$ in the equatorial plane, where we assume $x_o > x_1$ is outside the position of the inner light ring(s). We then place the celestial sphere at a radial position behind the observer, i.e. $x_1< x_o < x_{cs}$.

%\begin{figure}[tb]
    %\centering
    % What should this image show?
    %\includegraphics[width=0.95\textwidth]{observer.png}
    %\caption{\label{fig:observer}Illustration of the observer and celestial sphere positions.}
%\end{figure}

Our conditions for multiple shadows can then be neatly formulated in terms of the impact parameter $\eta$. The first condition is dictated by the spacetime itself. We require that the black hole be accessible from the observer's position, i.e. the throat at the inner light ring(s) must be open. This corresponds to trajectories satisfying
\begin{equation}
-\eta_1\leqslant \eta\leqslant \eta_1, \label{eq:eta1}
\end{equation}
where $\eta_1$ is the impact parameter for the inner light ring(s) as introduced in Section \ref{sec:photonorbits}.

The second condition derives from the observer's position. Note that for any observer, there will be a maximal impact parameter $\eta_o>0$ associated to their position. Any light ray seen by the observer, independent of its direction, must satisfy $\eta \leqslant \eta_o$. By the symmetry of the solution, the same value also forms a lower bound, so that the impact parameter $\eta$ for any light ray that can reach the observer satisfies
\begin{equation}
-\eta_o\leqslant \eta\leqslant \eta_o. \label{eq:etao}
\end{equation}

\begin{figure}[tb]
    \centering
    \includegraphics[width=0.45\textwidth]{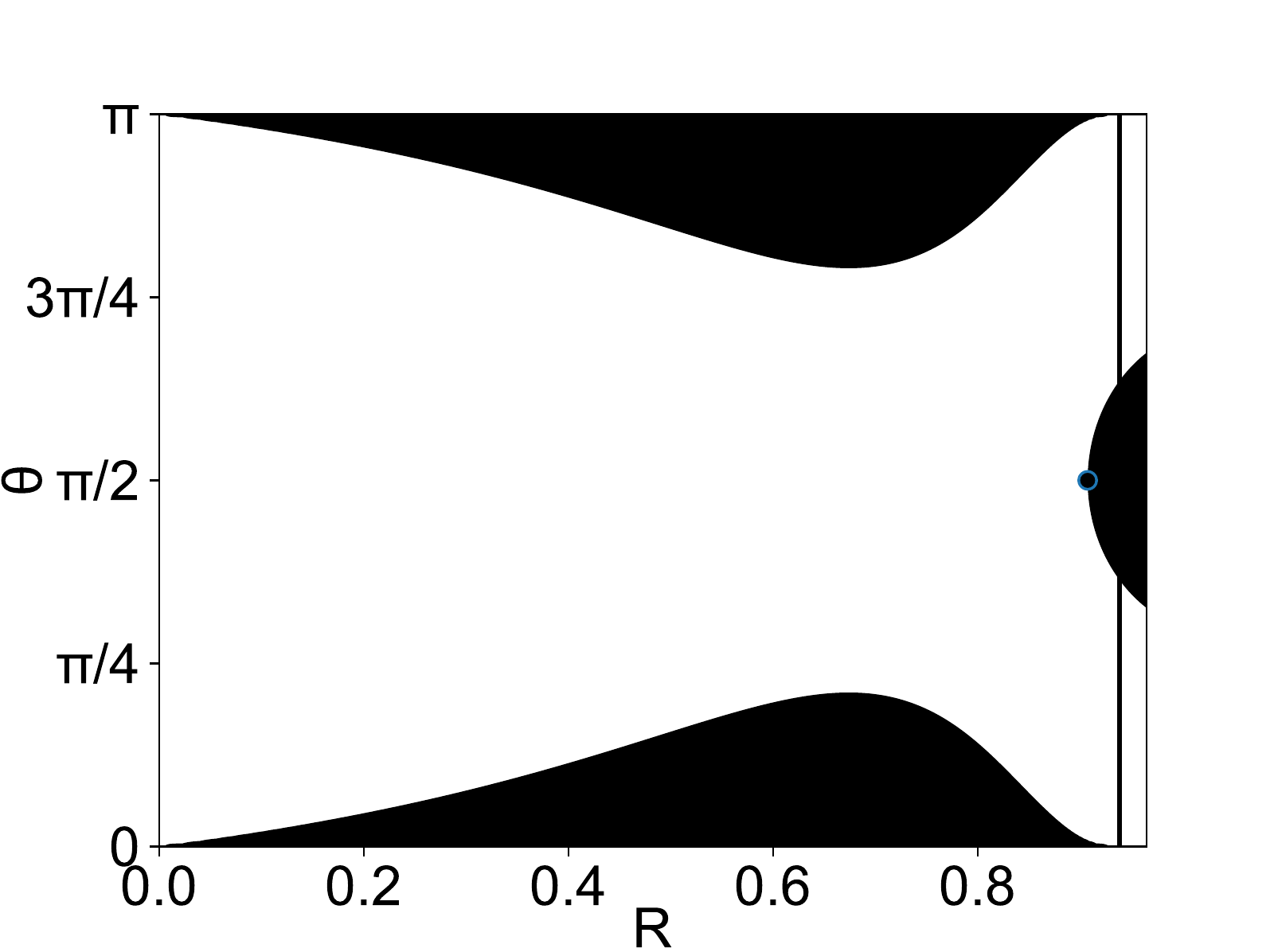}
    \caption{\label{fig:etao}The effective potential at impact parameter $\eta_o$ for an illustrative choice of observer (dot) and celestial sphere (vertical line). At this value of the impact parameter the observer is on the boundary of a forbidden region.}
\end{figure}

In terms of the potential, the value of $\eta_o$ corresponds to the threshold at which the observer is placed on the border of the allowed region of the spacetime, i.e. $\eta_o = h(x_o,y_o=0)$. An example of this is depicted in Fig. \ref{fig:etao}.

Finally, we have a third condition related to the location of the celestial sphere. Multiple shadows can form when photon trajectories scatter off the external forbidden region before plunging behind the horizon. Since the celestial sphere acts, effectively, as a radial cut-off for the spacetime, it can screen this dynamic for any trajectory with an impact factor that does not satisfy
\begin{equation}
|\eta| \geqslant \eta_{cs}, \label{eq:etacs}
\end{equation}
where $\eta_{cs} = h(x_{cs},0)$ is the impact parameter at which the external forbidden region just intersects the celestial sphere. To understand this condition, note that, as shown in Figs. \ref{fig:potentials1} and \ref{fig:potentials2}, the external forbidden region moves radially outward with decreasing $|\eta|$. Hence if, for some trajectory, $|\eta|$ falls below the limit  (\ref{eq:etacs}), the external forbidden region will withdraw completely behind the celestial sphere and be inaccessible.  

% Alex: this is the mathematical justification that the external region behaves as we claim in the part of the potential we're considering.
% If we want to include it in an appendix, it should probably be written up better.
\comment{
% Note: this is still written for the old h
To see this, consider $\eta = \pm\sqrt{h(x,\pi/2)}$ and hence $\frac{d\eta}{dx} = \frac{\frac{dh}{dx}}{\pm\sqrt{h(x,\pi/2)}}$. Thus $\frac{dx}{d|\eta|}=\frac{1}{\frac{d|\eta|}{dx}}<0$ iff $\frac{dh}{dx}<0$.
$$
\frac{dh}{dx}=\left[3-\frac{x+1}{x-1}+4a_2x(x+1)\right]\frac{(x+1)^2}{(x-1)}\exp(2a_a(x^2+1))
$$
is negative for $x\rightarrow\infty$. Recall that by our setup $x_{cs}>x_1$ since the celestial sphere is outside all the light rings. Recall further that at $x_1$ we either have $\left.\frac{dh}{dx}\right|_{x_1}=0$ (equatorial case), or there are no points along the equator where $\frac{dh}{dx}=0$. Therefore $\left.\frac{dh}{dx}\right|_{x_{cs}}<0$.
}

\begin{figure*}[tb]
    \centering
    \includegraphics[width=0.45\textwidth]{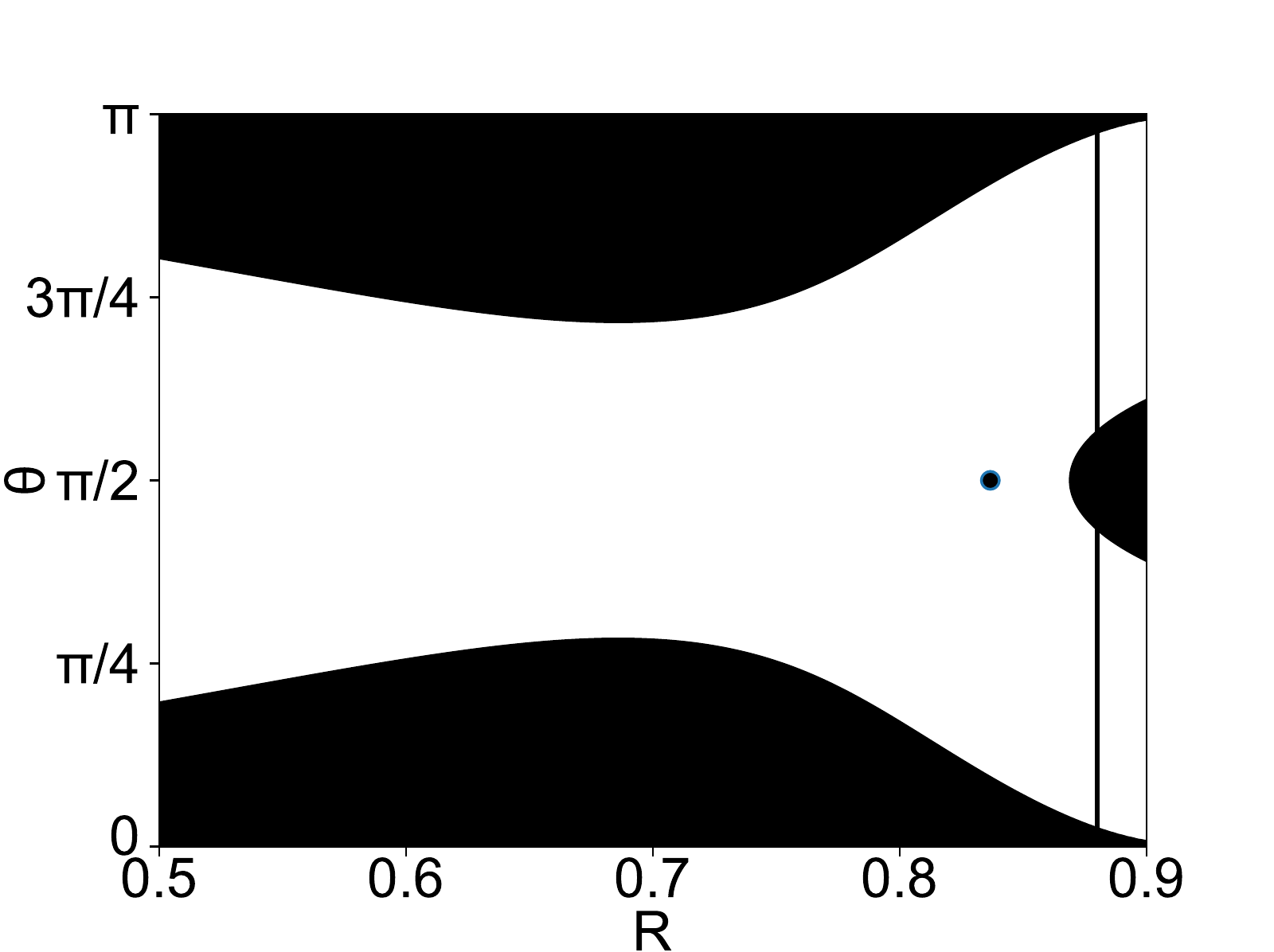}
    \includegraphics[width=0.45\textwidth]{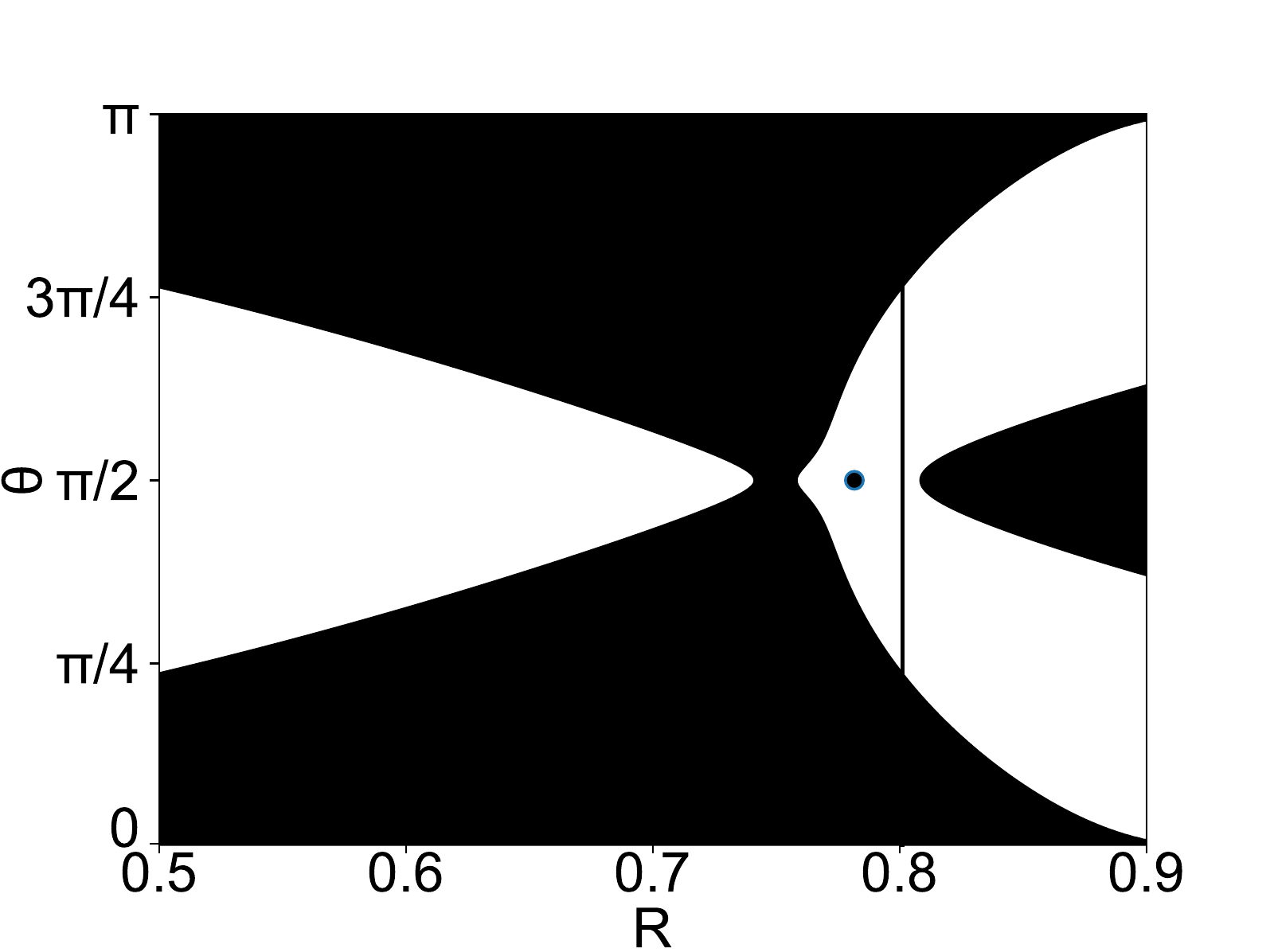}

    \caption{\label{fig:conditions}Qualitative features of the potential allowing multiple shadows (left), and not allowing multiple shadows (right). The observer position is marked by the dot, and the celestial sphere by the vertical line.}
\end{figure*}

Multiple shadows occur for impact parameters meeting all three conditions given in (\ref{eq:eta1})-(\ref{eq:etacs}). Letting $\bar\eta=\min(\eta_1,\eta_o)$, they can be expressed in condensed form as
% In interval form
%\begin{equation}
%\eta \in [-\bar\eta, \bar\eta] \setminus (-\eta_{cs}, %\eta_{cs}). \label{eq:interval}
%\end{equation}
% In inequalities (for me still less transparent, but maybe not that bad)
\begin{equation}
\eta_{cs} \leqslant |\eta| \leqslant \bar\eta
\label{eq:interval}
\end{equation}
We remark that this condition captures all possible cases of $a_2<0$ within the setup described here, including off-equatorial light rings. An immediate consequence of condition (\ref{eq:interval}) is that multiple shadows will not be possible, at least through the dynamic discussed here, on a given spacetime if we place our celestial sphere such that $\eta_{cs} > \bar\eta$. This can be intuitively understood by looking at the two potential plots shown in Fig. \ref{fig:conditions}. In the right panel we have an example with $\eta_{cs}>\eta_1$, in which case the throat at the inner light ring closes before the external forbidden region is accessible. By contrast, in the left panel we have an example with $\eta_{cs}<\eta_1$, so that for some trajectories both the external forbidden region and the black hole will be accessible, allowing for multiple shadows. 

\begin{figure*}[tb]
    \centering
    \includegraphics[width=0.95\textwidth,trim={81mm 45.5mm 94.5mm 48.3mm},clip]{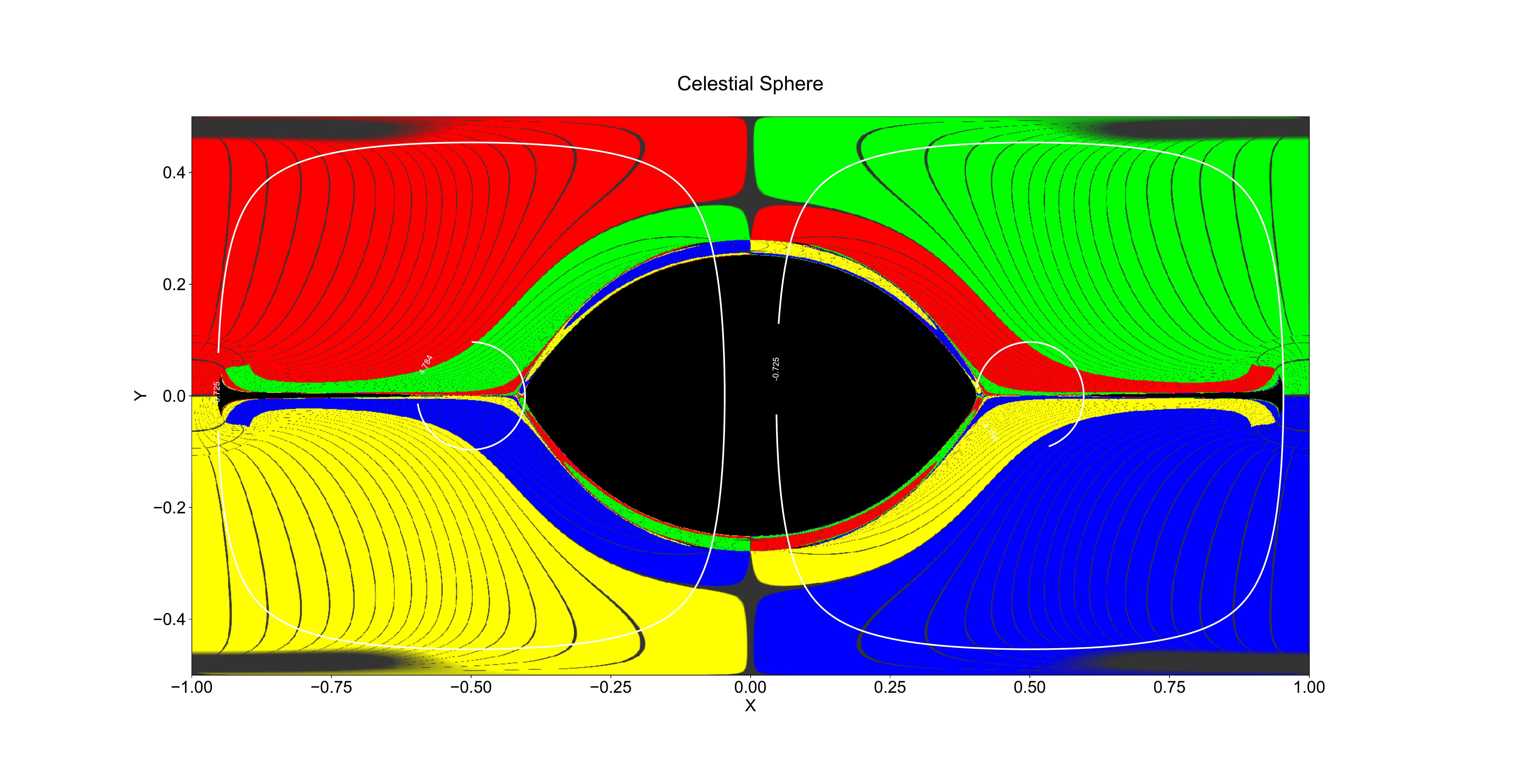}

    \caption{\label{fig:lensing1}Lensing image for quadrupole moment $a_2=-0.015$, an equatorial observer at $x_{o}=4.5$ and a celestial sphere at $x_{cs}=14.4$. Lines of constant impact parameter $\eta$ are shown for $\pm\eta_1\approx \pm4.784$ (small circles) and $\pm\eta_{cs}\approx \pm0.725$ (large circles).}
\end{figure*}

\begin{figure*}[tb]
    \centering
    \begin{subfigure}[t]{0.9\textwidth}
        \centering
        \includegraphics[width=\textwidth]{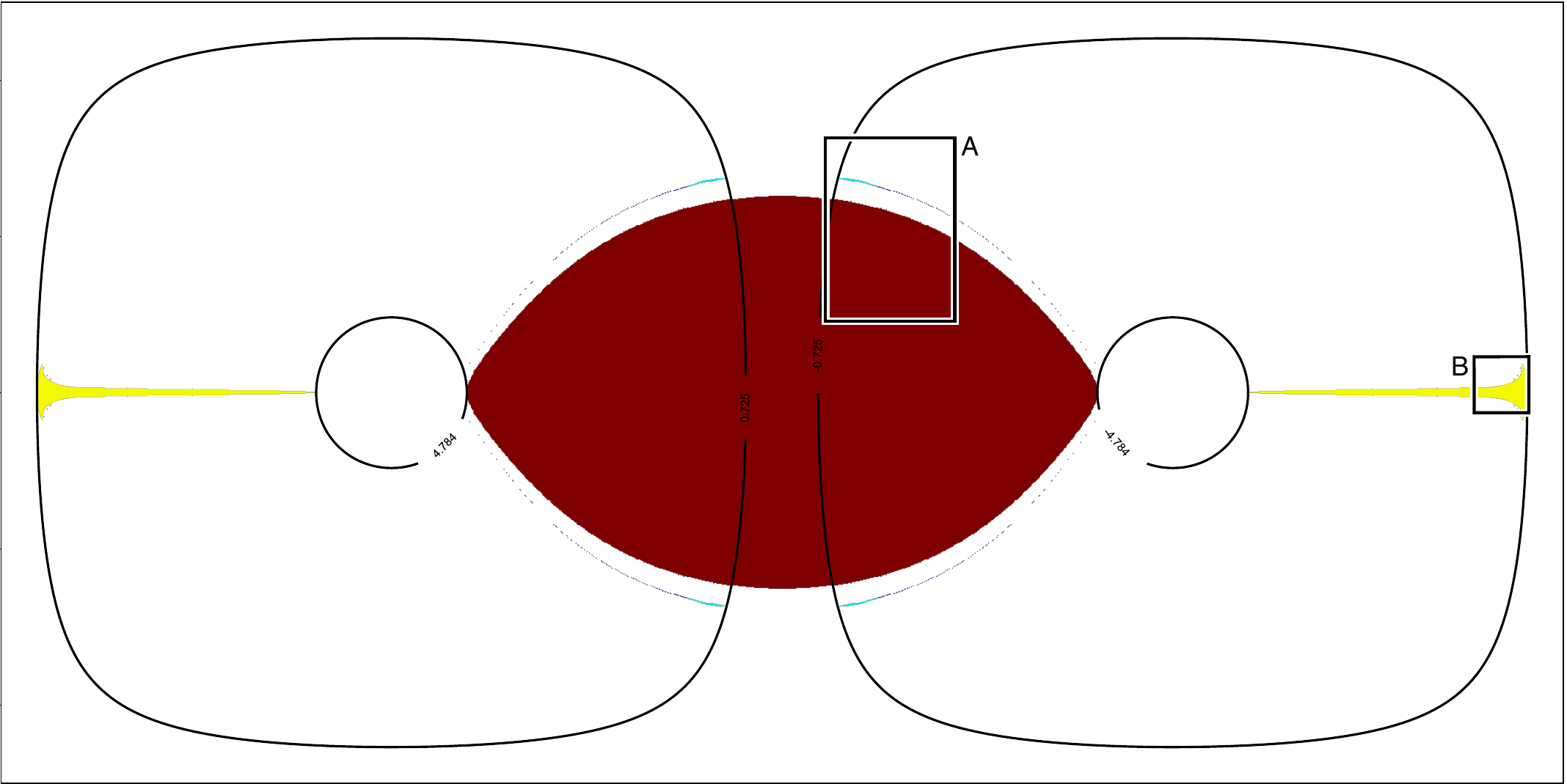}
        \caption{Shadow map with indicated regions of interest}
    \end{subfigure}
    %\\
    %\includegraphics[width=0.6\textwidth,trim={42mm 49mm 49mm  51mm},clip]{figures/equatorial/bounce-positions.pdf}
    %
    %\includegraphics[width=0.38\textwidth,trim={150mm 8mm 48mm 27mm},clip]{figures/equatorial/bounce-positions13.pdf}
    %\includegraphics[width=0.541\textwidth,trim={88mm 8mm 48mm 27mm},clip]{figures/equatorial/bounce-positions24.pdf}

    \begin{subfigure}[t]{0.35\textwidth}
        \centering
        \includegraphics[width=\textwidth,trim={351mm 36mm 95mm 39mm},clip]{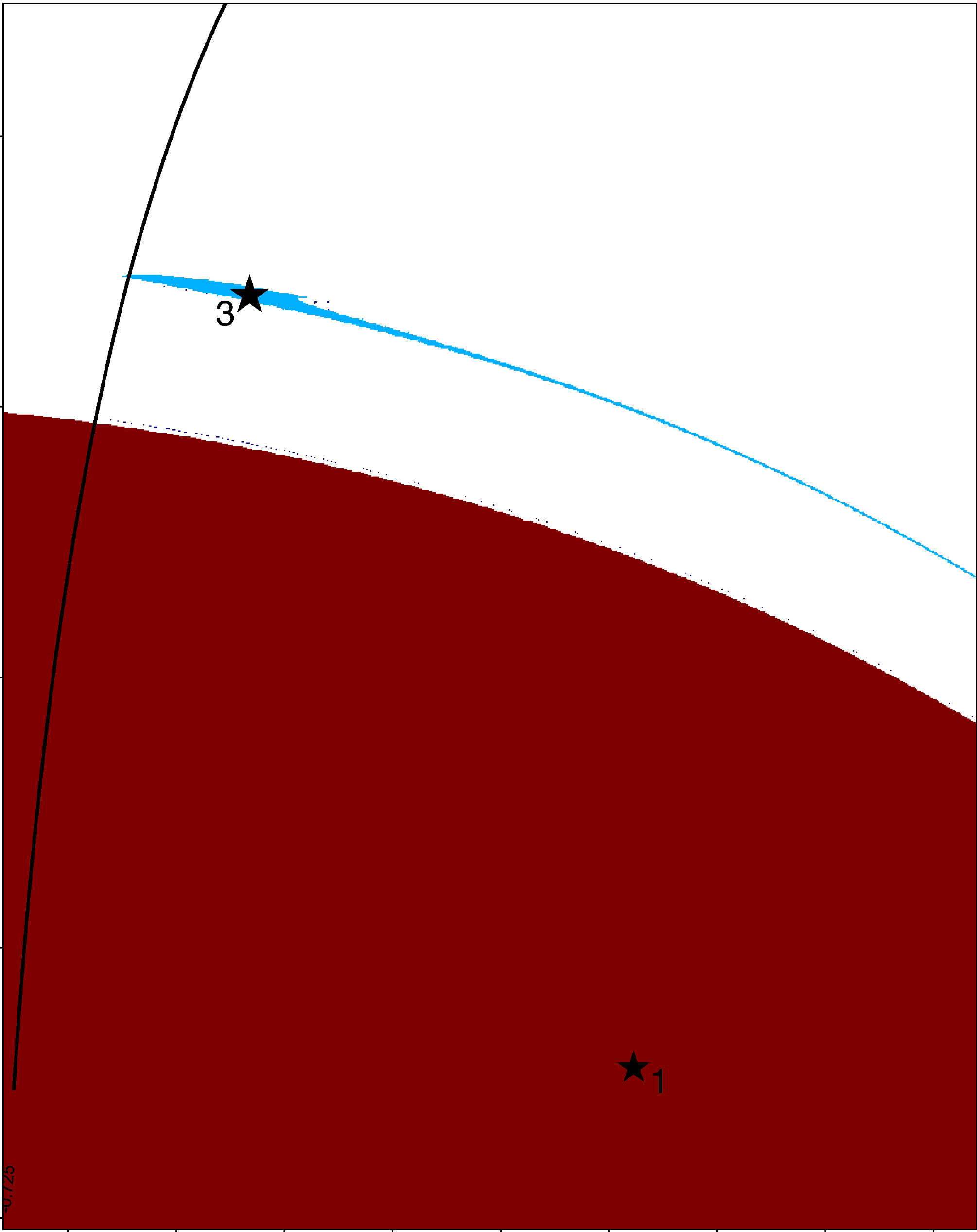}
        \caption{Region A, points 1 and 3 marked}
    \end{subfigure}
    \hspace{0.67cm}
    \begin{subfigure}[t]{0.5\textwidth}
        \centering
        \includegraphics[width=\textwidth,trim={264mm 36mm 95mm 39mm},clip]{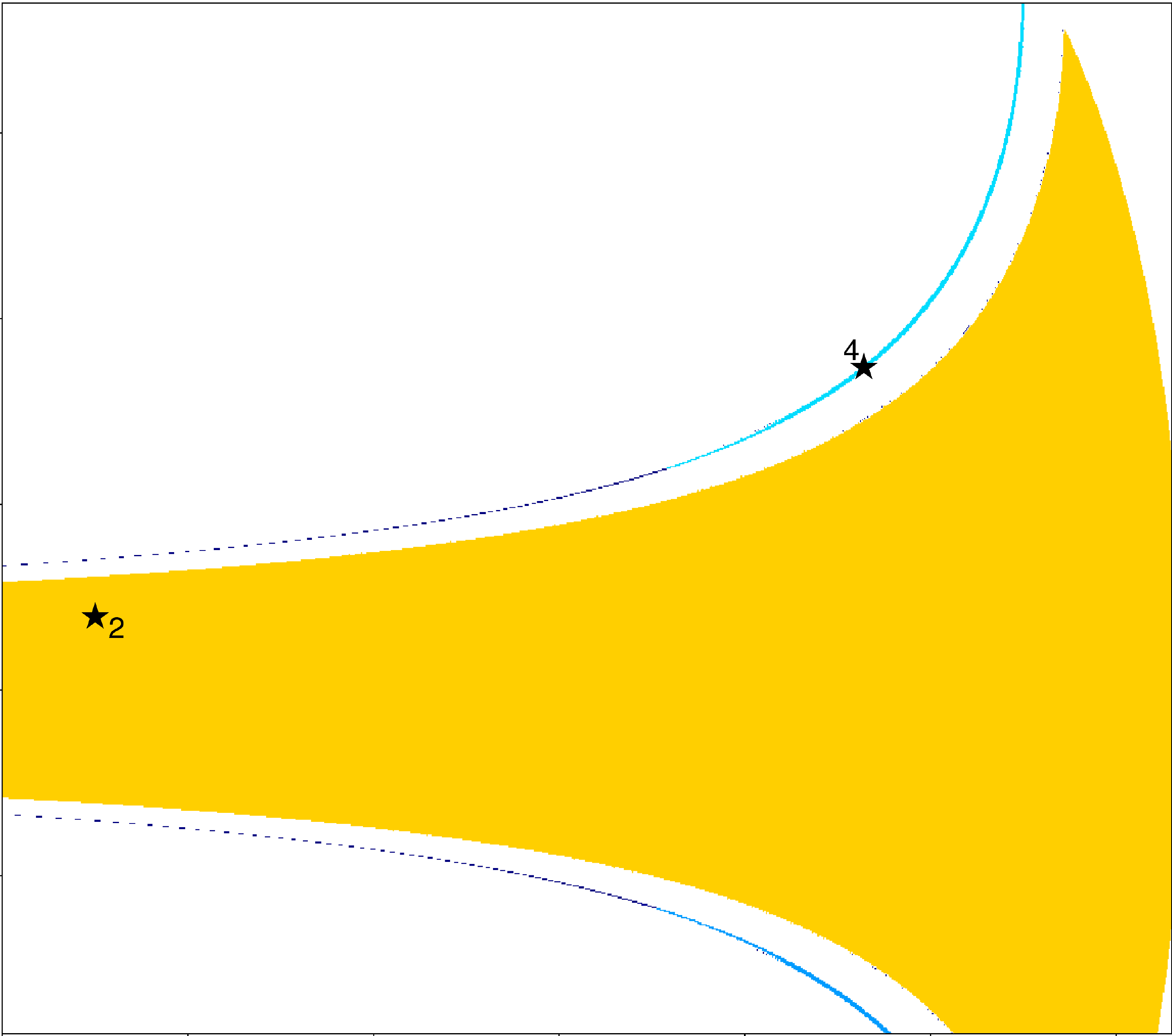}
        \caption{Region B, points 2 and 4 marked}
    \end{subfigure}

    \caption{\label{fig:shadows1}Shadow maps for quadrupole moment $a_2=-0.015$, an equatorial observer at $x_{o}=4.5$ and a celestial sphere at $x_{cs}=14.4$. Lines of constant impact parameter $\eta$ are shown for $\pm\eta_1\approx \pm4.784$ (small circles) and $\pm\eta_{cs}\approx \pm0.725$ (large circles).}
\end{figure*}

Finally, we illustrate the lensing image and shadow map for a spacetime with $a_2=-0.015>a_2^{crit}$ in Figs. \ref{fig:lensing1} and \ref{fig:shadows1} respectively. For these, the observer is placed at $(x_o,y_o) = (4.5,0)$ and the celestial sphere at $x_{cs} = 14.4$.

We find that additional shadows appear in Figs. \ref{fig:lensing1}-\ref{fig:shadows1} for $0.725=\eta_{cs} \leqslant \eta \leqslant \eta_1=4.784$ and symmetrically, $-\eta_1 < \eta < -\eta_{cs}$, as predicted by (\ref{eq:interval}). There are two main classes of shadow structure, corresponding on the one hand to trajectories that are initially directed towards the black hole and on the other to those that are directed away. Within each class there is a principal shadow surrounded by smaller additional shadows, or eyebrows, in a fractal structure; this can be seen in Fig. \ref{fig:shadows1}, up to the resolution of the image. 

\begin{figure*}[tb]
    \centering
    \begin{subfigure}[t]{0.45\textwidth}
        \centering
        \includegraphics[width=\textwidth,clip]{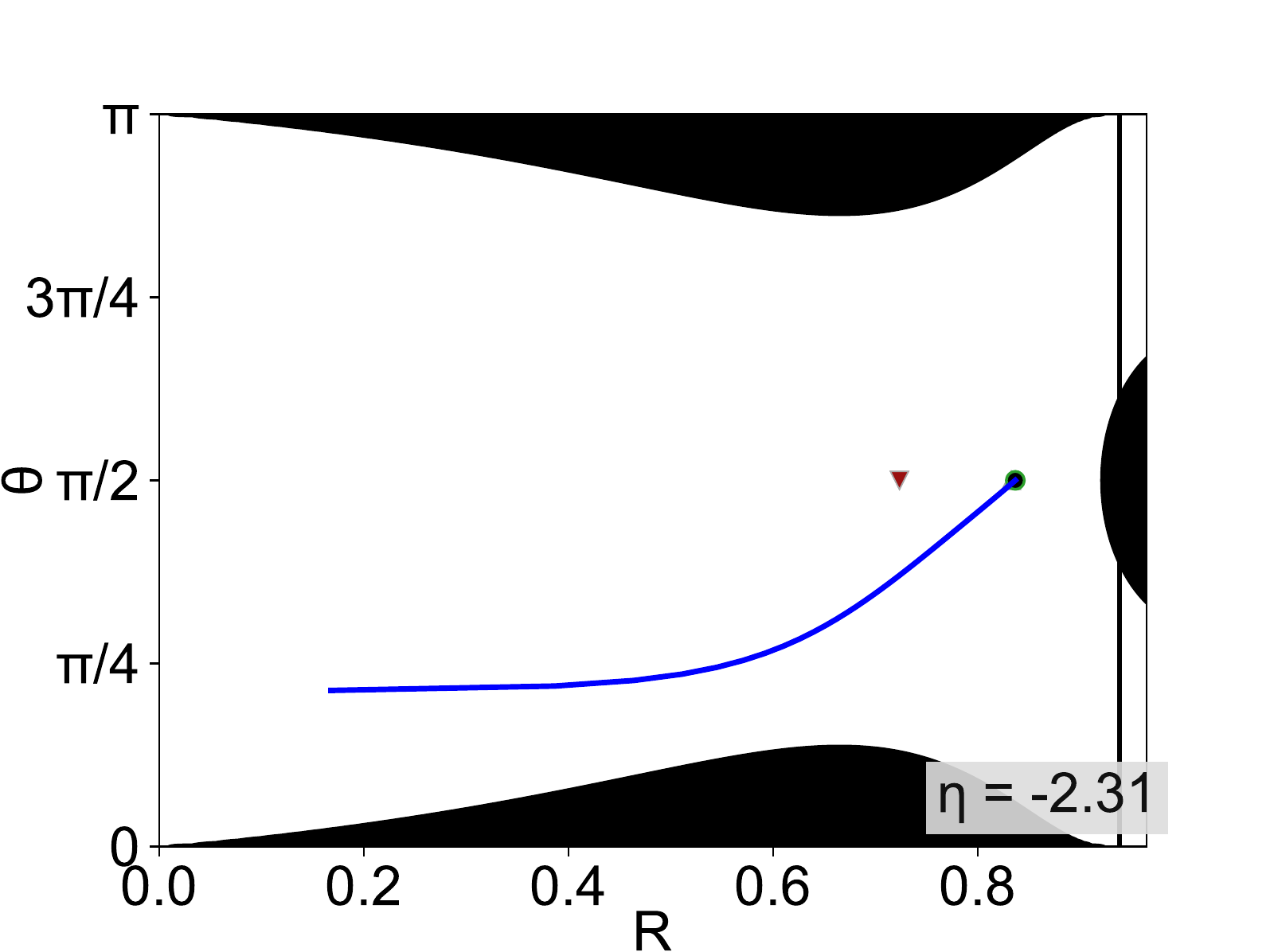}
        \caption{Trajectory corresponding to point 1}
    \end{subfigure}
    \begin{subfigure}[t]{0.45\textwidth}
        \centering
        \includegraphics[width=\textwidth,clip]{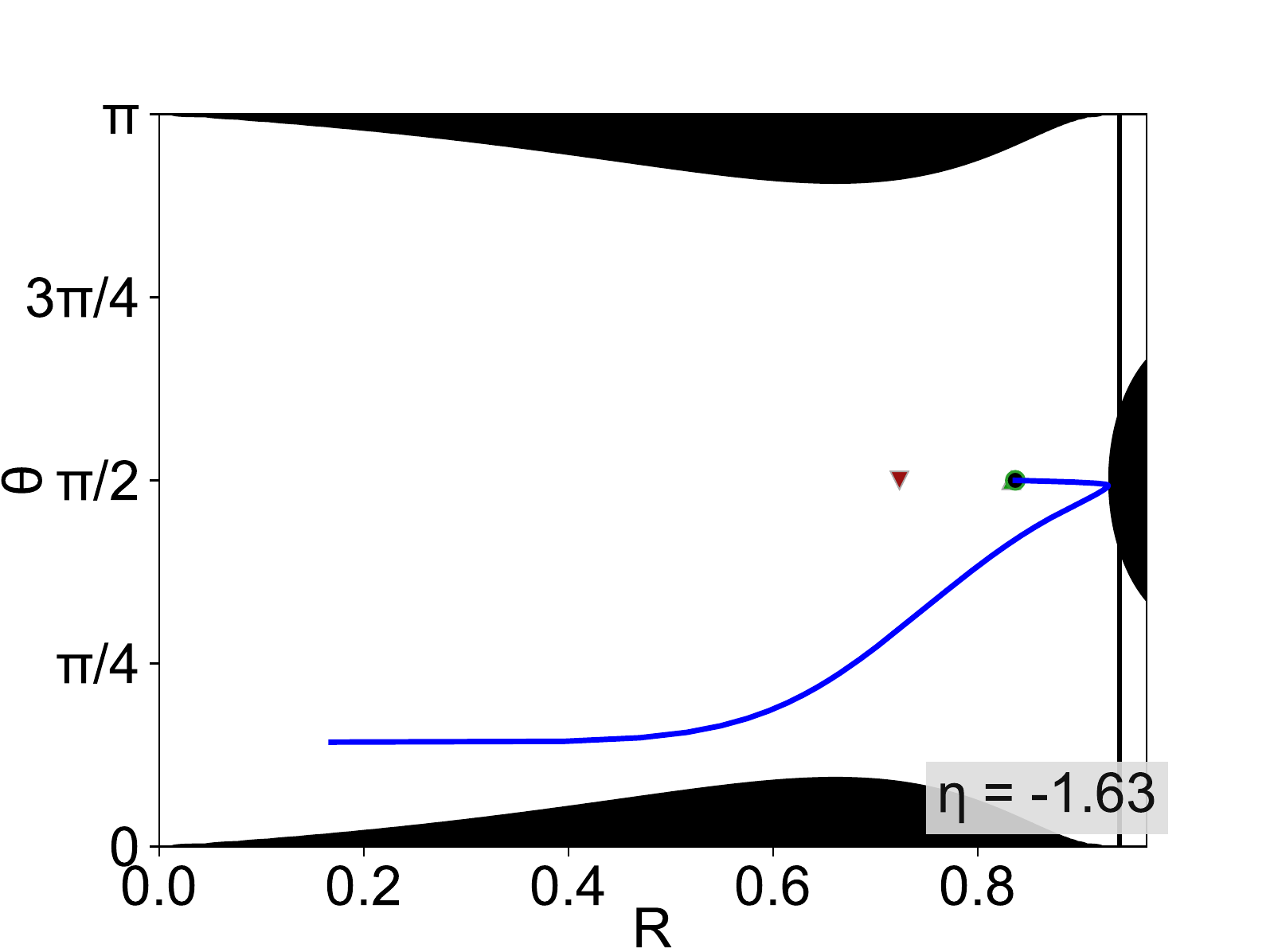}
        \caption{Trajectory corresponding to point 2}
    \end{subfigure}

    \begin{subfigure}[t]{0.45\textwidth}
        \centering
        \includegraphics[width=\textwidth,clip]{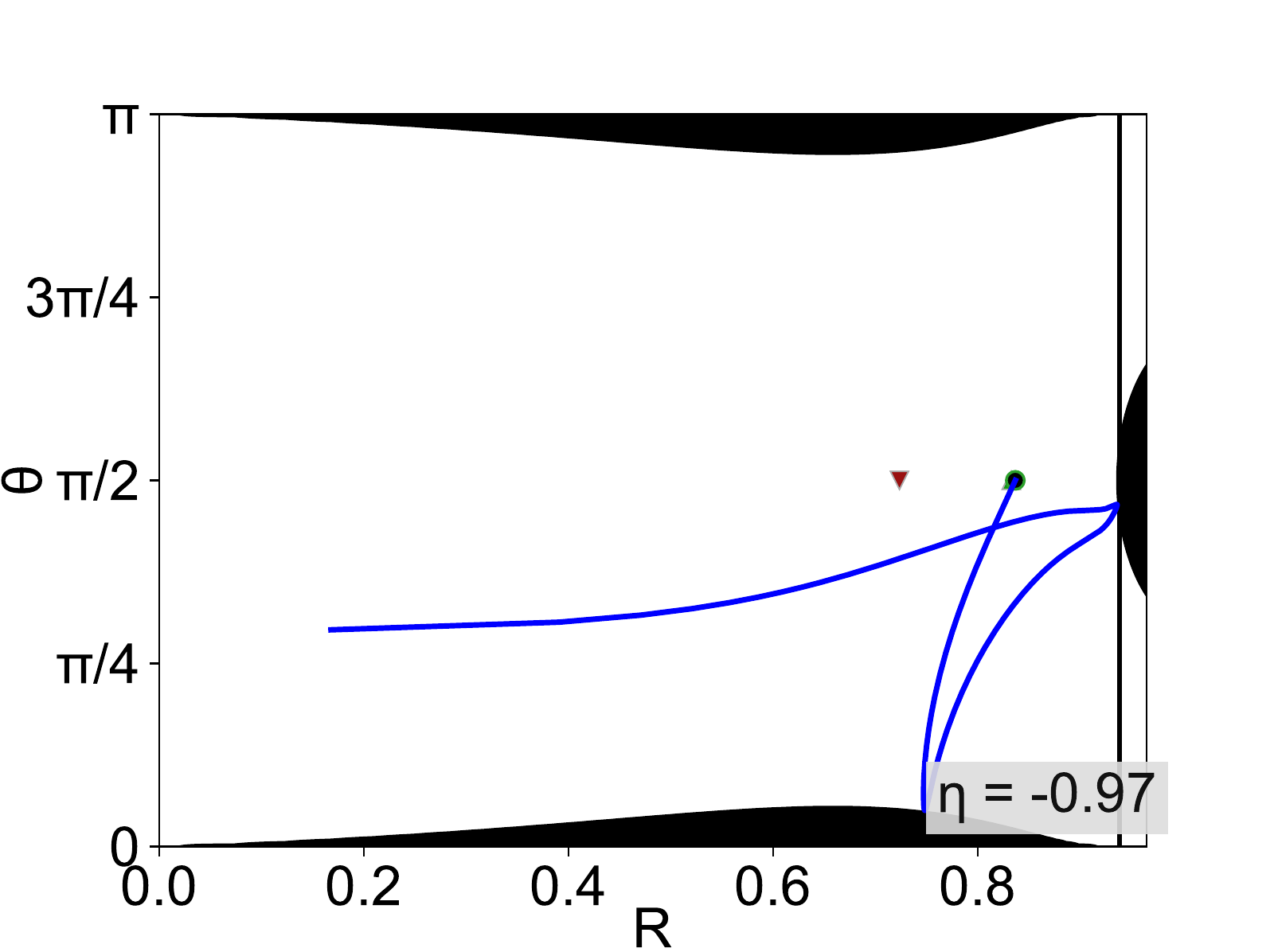}
        \caption{Trajectory corresponding to point 3}
    \end{subfigure}
    \begin{subfigure}[t]{0.45\textwidth}
        \centering
        \includegraphics[width=\textwidth,clip]{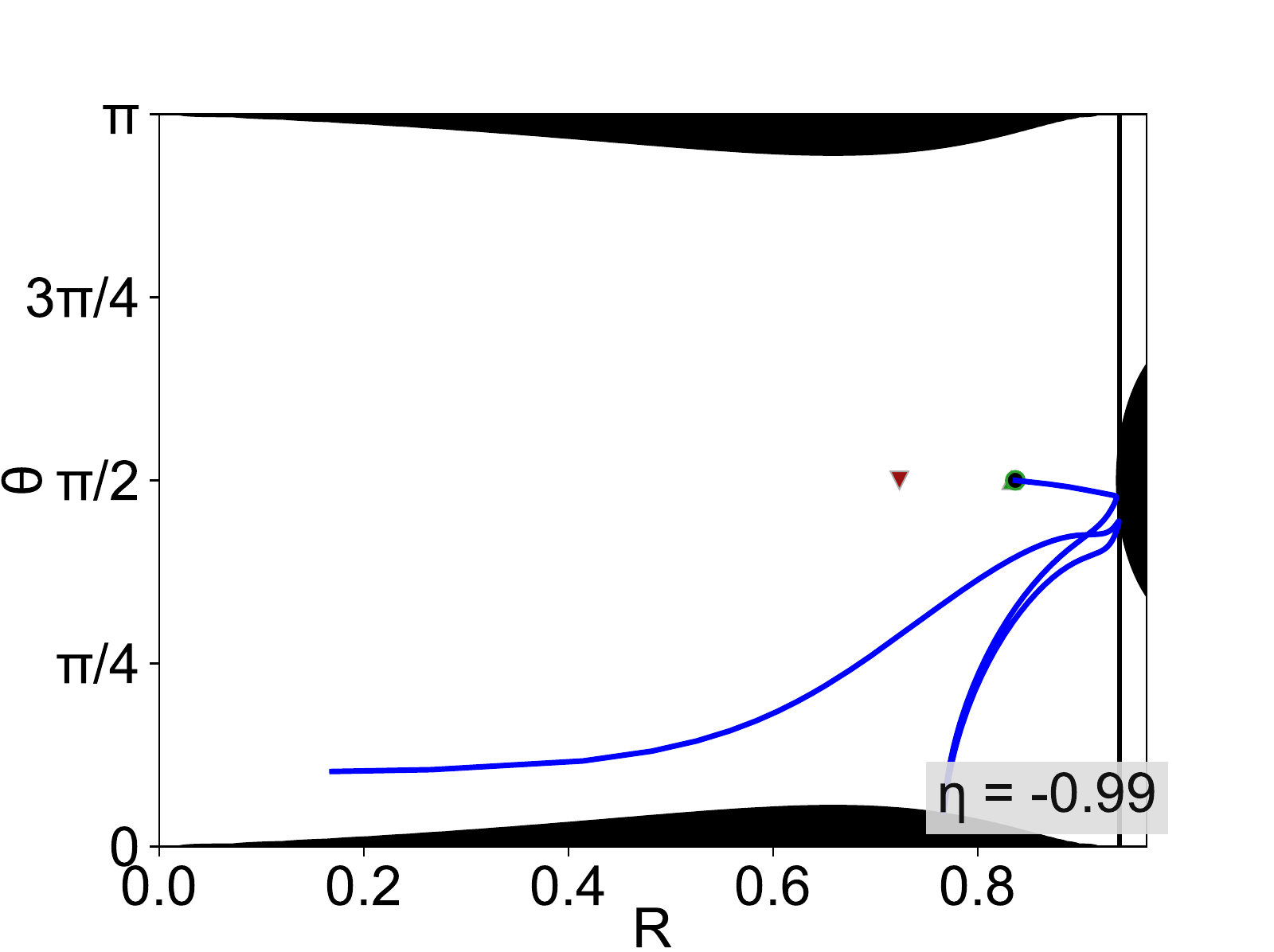}
        \caption{Trajectory corresponding to point 4}
    \end{subfigure}

    \caption{\label{fig:trajectories1}Photon trajectories for quadrupole moment $a_2=-0.015$, an equatorial observer (dot) at $x_{o}=4.5$ and a celestial sphere (vertical line) at $x_{cs}=14.4$ for different initial conditions, points 1-4, in Fig. \ref{fig:shadows1}. The position of the inner light ring is indicated by a red triangle, that of the outer light ring by a green triangle (behind the observer).}
\end{figure*}

Trajectories corresponding to the shadows are shown in the effective potential plots Fig. \ref{fig:trajectories1} for selected points in the shadow map Fig. \ref{fig:shadows1}. For trajectories directed towards the black hole, the principal shadow is composed of trajectories that plunge directly, with no radial turning points. This is depicted in Fig.\ref{fig:trajectories1}(a) for point 1, region A, in Fig. \ref{fig:shadows1}(b). Additional shadows in this class possess an even number of radial turning points and always scatter off the external forbidden region before plunging. This is depicted in Fig.\ref{fig:trajectories1}(c) for point 3, region A, in Fig. \ref{fig:shadows1}(b). Note that one can discern pixel sized shadows around these eyebrows corresponding (though not shown here) to trajectories with a still larger number of radial turning points. 

For trajectories that are initially directed away from the black hole, the principal shadow is composed of trajectories that scatter once off the external forbidden region before plunging. We illustrate this in Fig.\ref{fig:trajectories1}(b) for point 2, region B, in Fig. \ref{fig:shadows1}(c). Additional shadows in this class will possess an odd number of radial turning points and we find they also scatter off the external forbidden region before plunging. This is depicted in Fig.\ref{fig:trajectories1}(d) for point 4, region B, in Fig. \ref{fig:shadows1}(c).

\begin{figure}[tb]
    \centering
    \includegraphics[width=0.45\textwidth,clip]{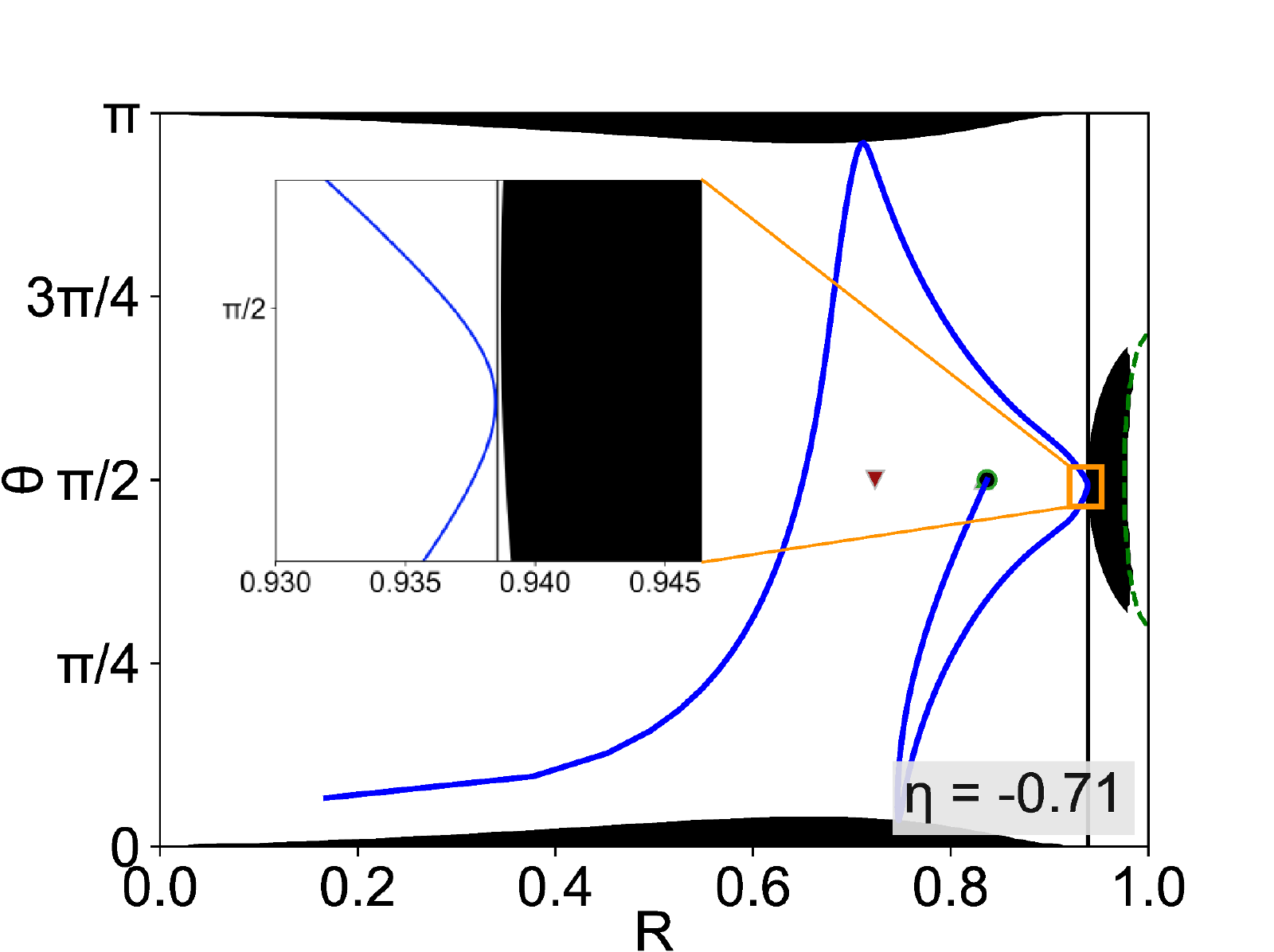}

    \caption{\label{fig:outofbounds} Photon trajectory for quadrupole moment $a_2=-0.015$, an equatorial observer (dot) at $x_{o}=4.5$ and a celestial sphere (vertical line) at $x_{cs}=14.4$ for a point in the shadow just left of the impact parameter $\eta_{cs}$ boundary in Fig. \ref{fig:shadows1}b. Inset shows zoom on the interaction with the external region without actually bouncing off of it.}
\end{figure}

However, it is important to emphasize that (\ref{eq:interval}) only provides a sufficient condition for multiple shadows; there are additional shadows that lie marginally \emph{outside} the bounds (\ref{eq:interval}). These can be seen on the left extremity of the secondary eyebrow (blue) in Fig. \ref{fig:shadows1}(b), just outside the $\eta_{cs}$ boundary line. A trajectory corresponding to this part of the eyebrow is shown in the effective potential plot Fig. \ref{fig:outofbounds}. It can be seen to exhibit the same qualitative feature of scattering from the external potential in the close vicinity of the external forbidden region.

\section{Conclusion}

In this paper we study the local shadow of the Schwarzschild black hole with a quadrupole distortion. In particular we discover the appearance of multiple shadows and provide a dynamical mechanism for their formation. The distorted solution describes the spacetime in the vicinity of the horizon of a static vacuum black hole interacting with an external matter distribution with the relevant symmetries. In this spacetime the photon motion is not expected to be integrable. It reduces to a 2-dimensional problem, so we can study qualitatively the dynamics by investigating a 2D effective potential, together with the properties of the light rings admitted by the spacetime. The influence of the external gravitational field modifies the light ring structure, and for negative quadrupole moments leads to the formation of a forbidden region for photon motion at large radial distances. We attribute the formation of multiple shadows to scattering by this forbidden region. Moreover, we isolate a sufficient condition under which this process can be realized, depending simultaneously on the quadrupole moment, and the position of the observer and the celestial sphere.

For negative quadrupole moments almost all the multiple shadow images, which we obtain numerically, result from such a scattering process, and satisfy the condition we propose. While we have no formal argument detailing other mechanisms for multiple shadow formation, our studies strongly suggest that the scattering mechanism we provide is the most prominent one. The analysis can be extended quite straightforwardly to the case of octupole distortion. Our preliminary investigations show that a similar mechanism is realized, caused by scattering from a certain potential barrier, which arises due to the influence of the external gravitational field. The conditions for observation of the multiple shadows also depend not only on the octupole moment, but also on the position of the observer and the celestial sphere.

\section*{Acknowledgments}
J.K. and P.N. gratefully acknowledge support by the DFG Research Training Group 1620 ``Models of Gravity''.  P.N. is partially supported by the Bulgarian NSF Grant $\textsl{DM 18/3}$.

\section*{Appendix: Stability of circular orbits and epicyclic frequencies}

In order to investigate the stability of the circular orbits in the equatorial plane we consider a small deviation from the circular motion $\tilde{x}^{\mu}(s) = x^{\mu}(s) + \xi^\mu(s)$, where $x^{\mu}(s)$ describes the circular orbit \footnote{A related calculation was also performed in \cite{Shoom:2017}.}. We substitute this expression in the geodesic equations, and considering terms up to linear order in $\xi^\mu(s)$, we obtain the following system, which describes the dynamics of the small perturbation \cite{Aliev:1981}, \cite{Aliev:1986}

\begin{eqnarray}\label{pert}
&&\frac{d^2\xi^\mu}{dt^2} + 2\gamma^\mu_\alpha\frac{d\xi^\alpha}{dt} + \xi^b\partial_b{\cal V}^{\mu} = 0\, , \quad b = x,y \nonumber \\[2mm]
&&\gamma^\mu_\alpha =\left[\Gamma^\mu_{\alpha\beta} u^\beta(u^0)^{-1}\right]_{y=0}\, , \nonumber \\[2mm]
&& {\cal V}^{\mu} = \left[\gamma^\mu_\alpha u^\alpha(u^0)^{-1}\right]_{y=0}.
\end{eqnarray}
The 4-velocity vector for the circular orbits in the equatorial plane is expressed as $u^\mu = u^0(1, 0, 0, \omega_0)$ by means of the orbital frequency $\omega_0=d\varphi/dt$, while $\Gamma^{\mu}_{\alpha\beta}$ are the Christoffel symbols. The equations for the $t$ and $\varphi$ components in $(\ref{pert})$ can be integrated leading to

\begin{eqnarray}
&&\frac{d\xi^A}{dt} + 2\gamma^A_x\xi^x = 0\, , \quad A = t,\varphi \nonumber \\[2mm]
&&\frac{d^2\xi^x}{dt^2} + \omega_x^2\xi^x = 0\, , \label{pertx} \\[2mm]
&&\frac{d^2\xi^y}{dt^2} + \omega_y^2\xi^y = 0\, ,  \label{perty} \\[2mm] \nonumber
&&\omega_x^2 = \partial_x {\cal V}^x - 4\gamma^x_A\gamma^A_x, \quad \omega_y^2 = \partial_y{\cal V}^y. \nonumber
\end{eqnarray}
Thus, we obtain two decoupled equations for the perturbations  in radial and axial direction. The dynamics of the perturbations is determined by the sign of the quantities $\omega_x^2$ and $\omega_y^2$. For positive values eqs. $(\ref{pertx})$-$(\ref{perty})$ describe a pair of harmonic oscillators, i.e.  small deviations from the circular orbits will oscillate in their vicinity with frequencies $\omega_x$ and $\omega_y$ in radial and vertical direction,  respectively. In this case the circular orbit is stable in linear approximation, and the quantities $\omega_x$ and $\omega_y$ are called epicyclic frequencies. If one of the quantities $\omega_x^2$ and $\omega_y^2$ is negative, the circular orbit is unstable, since small perturbations in the corresponding direction will deviate exponentially from it.

Performing the calculations for the distorted Schwarzschild solution,  we obtain the following expressions for the epicyclic frequencies in the  case of null geodesics

\begin{eqnarray}\label{omega_null}
\omega^2_y= -\omega^2_x = \frac{\omega_0^2e^{-2\gamma}}{1-a_2x(x^2-1)}(1+2a_2x^3),
\end{eqnarray}
where $\omega_0$ is the orbital frequency. From this expression it is obvious that the regions of stability with respect to radial and vertical perturbations are complementary, i.e. an orbit which is stable with respect to radial perturbations will be unstable with respect to vertical ones.  The domain of existence of circular orbits, and the regions of stability with respect to radial and vertical perturbations are illustrated in Fig. \ref{st}. The curve $\omega^2_y = \omega^2_x =0$ defined by eq. ($\ref{omega_null}$) is presented in blue. The region under the blue curve corresponds to positive values of $\omega^2_x$, and consequently negative values of $\omega^2_y$, while in the region above the blue curve $\omega^2_x<0$ and $\omega^2_y>0$ are satisfied. It is seen that for positive values of the multipole moment $a_2$ all the circular orbits are stable with respect to vertical perturbations, however unstable with respect to radial ones.

Using eq. ($\ref{LR0}$) the epicyclic and the orbital frequencies can be further expressed as functions only of the position of the circular orbit $x_1$

\begin{eqnarray}
\omega^2_y &=& -\omega^2_x = -\frac{2\omega_0^2\,e^{-2\gamma}}{x_1(x_1^2-1)}(x_1^3 -3x_1^2 +1)\, , \\[2mm]
\omega^2_0 &=& \frac{x_1-1}{(x_1+1)^3}\exp\left[\frac{(x_1-2)(x_1^2+1)}{x_1(x_1^2-1)}\right]\, ,  \nonumber \\[2mm]
\gamma&=&\frac{x_1-2}{16x_1^2(x_1^2-1)}(x_1^3+14x_1^2-x_1+2)\,. \nonumber
\end{eqnarray}


\begin{thebibliography}{tbds}

\bibitem{Doeleman}
S. Doeleman, E. Agol, D. Backer et al., ``Imaging an event horizon: submm-VLBI of a super massive black hole'', Astro2010: The Astronomy and Astrophysics Decadal Survey, Science White Papers, No. 68 (2009).

\bibitem{Wilson}
C. Wilson-Hodge, P. Ray, K. Gendreau et al., ``STROBE-X: X-ray timing: Spectroscopy on dynamical timescales from microseconds to years'', in American Astronomical Society Meeting Abstracts, Vol. 229, American Astronomical Society Meeting Abstracts, 309.04 (2017).

\bibitem{Feroci}
M. Feroci, E. Bozzo, S. Brandt et al., ``The LOFT mission concept: a status update'', in Proc. SPIE, Vol. 9905, Society of Photo-Optical Instrumentation Engineers (SPIE) Conference Series (2016) 99051R.

\bibitem{Zhang}
S. Zhang, M. Feroci, A. Santangelo et al., ``eXTP: enhanced X-ray timing and polarimetry mission'', in Proc. SPIE, Vol. 9905, Space Telescopes and Instrumentation 2016: Ultraviolet to Gamma Ray (2016) 99051Q.

\bibitem{Nedkova:2013}
P. Nedkova, V. Tinchev, S. Yazadjiev, 
``Shadow of a rotating traversable wormhole'',
Phys. Rev. D88 (2013) 124019.

\bibitem{Sakai}
N. Sakai, H. Saida, and T. Tamaki,
``Gravastar shadows'',
Phys. Rev. D 90 (2014) 104013.

\bibitem{Bardeen}
J. M. Bardeen, 
``Timelike and null geodesics in the Kerr metric,'' in {\it Black Holes}, editors Dewitt, C. and Dewitt, B. S., pp. 215-239, Gordon and Breach, New York (1973).

\bibitem{Grenzebach:2014}
A. Grenzebach, V. Perlick and L\"{a}mmerzahl, 
``Photon regions and shadows of Kerr-Newman-NUT black holes with a cosmological constant'',
Phys. Rev. D89 (2014) 124004.

\bibitem{Grenzebach:2015a}
A. Grenzebach,  
``Aberrational effects for shadows of black holes'',
 Fund. Theor. Phys. 179 (2015) 823.
 
\bibitem{Grenzebach:2015b}
A. Grenzebach, V. Perlick and L\"{a}mmerzahl, 
``Photon regions and shadows of accelerated black holes '',
Int. J. Mod. Phys. D24 (2015) 1542024.

\bibitem{Rezzolla:2015}
 A. Abdujabbarov, L. Rezzolla, B.  Ahmedov, ``A coordinate independent characterisation of a black-hole shadow'',
 Mon. Not. Roy. Astron. Soc. 454 (2015) 2423.

\bibitem{Tinchev:2014}
V. Tinchev, S. Yazadjiev,
``Possible imprints of cosmic strings in the shadows of galactic black holes'',
Int. J. Mod. Phys. D23 (2014) 1450060.


\bibitem{Amarilla:2010}
L. Amarilla, E. F. Eiroa and G. Giribet, 
``Null geodesics and shadow of a rotating black hole in extended Chern-Simons modified gravity'',
Phys. Rev. D 81 (2010) 124045.

\bibitem{Amarilla:2012}
L. Amarilla and E. F. Eiroa, 
``Shadow of a rotating braneworld black hole,''
Phys. Rev. D 85 (2012) 064019.

\bibitem{Amarilla:2013}
L. Amarilla and E. F. Eiroa  
``Shadow of a Kaluza-Klein rotating dilaton black hole '',
Phys.Rev. D87 (2013)  044057.

\bibitem{Cunha:2016a}
P. V.P. Cunha, C. Herdeiro, B. Kleihaus, J. Kunz, E. Radu
``Shadows of Einstein-dilaton-Gauss-Bonnet black holes'',
Phys.Lett. B768 (2017) 373.

\bibitem{Yazadjiev:2018}
T. Vetsov, G. Gyulchev, S. Yazadjiev, 
``Shadows of black holes in vector-tensor Galileons modified gravity,''
 arXiv:1801.04592


\bibitem{Tsukamoto}
N. Tsukamoto, Z. Li and C. Bambi, 
``Constraining the spin and the deformation parameters from the black hole shadow'',
JCAP 1406 (2014) 043.

\bibitem{Takahashi}
R. Takahashi, 
``Shapes and positions of black hole shadows in accretion disks and spin parameters of black holes'',
Astrophys.J. 611 (2004) 996. 

\bibitem{Hioki}
K. Hioki, K. Maeda, 
``Measurement of the Kerr spin parameter by observation of a compact object's shadow'',
Phys. Rev. D 80 (2009) 024042.

\bibitem{Li}
Z. Li, C. Bambi
``Measuring the Kerr spin parameter of regular black holes from their shadow'',
JCAP 1401 (2014) 041.


\bibitem{Johannsen}
T. Johannsen,
``Testing the no-hair theorem with observations of black holes in the electromagnetic spectrum'', 
Class. Quant. Grav. 33 (2016) 124001.

\bibitem{Liu}
K. Liu, N. Wex, M. Kramer, J. M. Cordes and T. J. W. Lazio, 
``Prospects for probing the spacetime of Sgr A* with pulsars'',
Astrophys. J. 747 (2012) 1.

\bibitem{Broderick}
A. E. Broderick , T. Johannsen, A. Loeb, D. Psaltis,
``Testing the no-hair theorem with event horizon telescope observations of Sagittarius A*'',
Astrophys.J. 784 (2014) 7.


\bibitem{Bambi:2015}
C. Bambi, 
``Testing the Kerr paradigm with the black hole shadow'', [arXiv:1507.05257].

\bibitem{Yumoto}
A. Yumoto, D. Nitta, T. Chiba, N. Sugiyama,
``Shadows of multi-black holes: analytic exploration'',
Phys. Rev. D 86 (2012) 103001.

\bibitem{Nitta}
 D. Nitta, T. Chiba, N. Sugiyama,
``Shadows of colliding black holes'',
 Phys.Rev. D84 (2011) 063008.

\bibitem{Shipley}
J. O. Shipley, S. Dolan, 
``Binary black hole shadows, chaotic scattering and the Cantor set'',
Class. Quant. Grav. 33 (2016) 175001.

\bibitem{Cunha:2015}
P. V. P. Cunha, C. A. R. Herdeiro, E. Radu, H. F. Runarsson,
``Shadows of Kerr black holes with scalar hair'', 
Phys. Rev. Lett. 115 (2015) 211102.

\bibitem{Cunha:2016}
P. V. P. Cunha, J. Grover, C. Herdeiro, E. Radu, H. Runarsson, A. Wittig,
``Chaotic lensing around boson stars and Kerr black holes with scalar hair'', 
Phys. Rev. D 94 (2016) 104023.

\bibitem{Bohn}
Bohn, et. al. , ``What does a binary black hole merger look like?'', Class. Quant. Grav. 32 (2015) 065002.

\bibitem{Geroch}
R. Geroch and J. B. Hartle,
``Distorted black holes,''
 J. Math. Phys. 23 (1982) 680.

\bibitem{Doroshkevich}
A. Doroshkevich, Ya. Zeldovich, I. Novikov,
``Gravitational collapse of nonsymmetric and rotating masses,''
Sov. Phys. JETP 22 (1966) 122.

\bibitem{Tomimatsu}
A. Tomimatsu,
``Distorted rotating black holes'',
Phys. Lett. A 103 (1984) 374.

\bibitem {Breton:1997}
N.~Bret\'{o}n, T.~Denisova, and V.~Manko,
``A Kerr black hole in the external gravitational field'',
Phys.\ Lett.\ A 230 (1997) 7.


\bibitem{Fairhurst}
 S. Fairhurst and B. Krishnan,
 ``Distorted black holes with charge'',
 Int. J. Mod. Phys. D 10 (2001) 691.

\bibitem{Yazadjiev}
S. S. Yazadjiev,
``Distorted charged dilaton black holes'',
Class. Quant. Grav. 18 (2001) 2105.

\bibitem {Breton:1998}
N. Bret\'{o}n, A. A. Garc\'{\i}a, V. S. Manko, and T. E. Denisova,
``Arbitrarily deformed Kerr Newman black hole in an external gravitational field'',
 Phys. Rev. D 57 (1998) 3382.

\bibitem{Abdolrahimi:2010}
S.~Abdolrahimi,  A.~Shoom, D.~Page,
``Distorted five-dimensional vacuum black hole'',
Phys.\ Rev.\  D 82 (2010) 124039.

\bibitem{Abdolrahimi:2013}
S.~Abdolrahimi, A.~Shoom,
``Distorted five-dimensional electrically charged black holes'',
Phys. Rev. D 89 (2014) 024040.

\bibitem{Nedkova:2014}
S.~Abdolrahimi, J.~Kunz, P.~Nedkova,
``Myers-Perry black hole in an external gravitational field'',
Phys.Rev. D91 (2015) 064068.

\bibitem{Kunz:2017}
J. Kunz, P. Nedkova, S. Yazadjiev, 
``Magnetized black holes in an external gravitational field'', 
Phys. Rev. D 96 (2017) 024017.


\bibitem{Abdolrahimi:2015a}
S.~Abdolrahimi, J.~Kunz, P.~Nedkova, C.~Tzounis
``Properties of the distorted Kerr black hole '',
JCAP 1512 (2015) 009.

\bibitem{Shoom:2016}
A. Shoom, C. Walsh, I. Booth, 
``Geodesic motion around a distorted static black hole'',
Phys.Rev. D93 (2016) 064019.

\bibitem{Shoom:2017}
A. Shoom, 
``Metamorphoses of a photon sphere'',
Phys.Rev. D96 (2017)  084056.


\bibitem{Abdolrahimi:2015b}
S. Abdolrahimi, R. Mann, C. Tzounis, 
``Distorted local shadows'',
Phys. Rev. D 91 (2015) 084052.

\bibitem{Abdolrahimi:2015c}
S. Abdolrahimi, R. Mann, C. Tzounis, 
``Double images from a single black hole'', 
Phys. Rev. D 92 (2015) 124011.

\bibitem{Grover}
J. Grover, A. Wittig,
``Black hole shadows and invariant phase space structures'',
Phys. Rev. D 96 (2017) 024045.

\bibitem{Cardoso:2009}
V. Cardoso, A. S. Miranda, E. Berti, H. Witek, and
V. T. Zanchin, 
``Geodesic stability, Lyapunov exponents and quasinormal modes'',
Phys. Rev. D 79, 064016 (2009).

\bibitem{Dolan:2016}
S. Dolan, J. Shipley
``Stable photon orbits in stationary axisymmetric electrovacuum spacetimes,''
Phys.Rev. D 94, 044038 (2016).

\bibitem{Cunha:2017}
P. V. P Cunha, E. Berti, C. A. R. Herdeiro,
'' Light-Ring Stability for Ultracompact Objects '',
Phys. Rev. Lett. 119 (2017) 04211.


\bibitem{Hajicek:1973}
P. Hajicek, 
``Exact models of charged black holes II. Axisymmetric stationary horizons'',
Commun. Math. Phys. 34 (1973) 53.


\bibitem{Aliev:1981}
A. N. Aliev and D. V. Gal’tsov,
``Radiation from relativistic particle in non-geodesic motion in a strong gravitational field'', 
Gen. Relat. Gravit. 13 (1981) 899.

\bibitem{Aliev:1986}
A. N. Aliev, D. V. Gal’tsov and V. I. Petukhov,
``Negative absorption near a magnetized black hole: Black hole masers'',
Astr. Space Sci. 124 (1986) 137.

\end{thebibliography}
\end{document}